\documentclass[sigconf]{acmart}
\usepackage{multirow}
\usepackage{graphicx}

\AtBeginDocument{%
  }

\copyrightyear{2025}
\acmYear{2025}
\setcopyright{cc}
\setcctype{by}
\acmConference[CHI '25]{CHI Conference on Human Factors in Computing Systems}{April 26-May 1, 2025}{Yokohama, Japan}
\acmBooktitle{CHI Conference on Human Factors in Computing Systems (CHI '25), April 26-May 1, 2025, Yokohama, Japan}\acmDOI{10.1145/3706598.3713722}
\acmISBN{979-8-4007-1394-1/25/04}

\begin{document}

\title{Reimagining Personal Data: Unlocking the Potential of AI-Generated Images in Personal Data Meaning-Making}

\author{Soobin Park}
\email{soobinpark@kaist.ac.kr}
\affiliation{%
  \institution{Department of Industrial Design, KAIST}
  \city{Daejeon}
  \country{Republic of Korea}
}

\author{Hankyung Kim}
\email{hkkim31@kaist.ac.kr}
\affiliation{%
  \institution{Department of Industrial Design, KAIST}
  \city{Daejeon}
  \country{Republic of Korea}
}

\author{Youn-kyung Lim}
\email{younlim@kaist.ac.kr}
\affiliation{%
  \institution{Department of Industrial Design, KAIST}
  \city{Daejeon}
  \country{Republic of Korea}
}

\begin{abstract}
Image-generative AI provides new opportunities to transform personal data into alternative visual forms. In this paper, we illustrate the potential of AI-generated images in facilitating meaningful engagement with personal data. In a formative autobiographical design study, we explored the design and use of AI-generated images derived from personal data. Informed by this study, we designed a web-based application as a probe that represents personal data through generative images utilizing Open AI’s GPT-4 model and DALL-E 3. We then conducted a 21-day diary study and interviews using the probe with 16 participants to investigate users’ in-depth experiences with images generated by AI in everyday lives. Our findings reveal new qualities of experiences in users’ engagement with data, highlighting how participants constructed personal meaning from their data through imagination and speculation on AI-generated images. We conclude by discussing the potential and concerns of leveraging image-generative AI for personal data meaning-making. 

\end{abstract}

\ccsdesc[500]{Human-centered computing~Empirical studies in HCI}

\keywords{generative AI, image generation, human-AI interaction, personal data, reflection, introspection, self-tracking
}

\maketitle

\section{Introduction}
With the advances in data-driven digital technologies, people can collect various types of data about themselves, i.e., personal data in their everyday lives, in both explicit and implicit ways. In the Human-Computer Interaction (HCI) research community, the importance of meaning-making from such personal data has been highlighted \cite{elsden2016quantified, cocskun2023data, mekler2019framework, sosa2018data,karyda2021data}, alongside ongoing attempts on more creative and alternative ways of experiencing personal data \cite{elsden2016fitter, desjardins2023making, lee2023data, gaver2007enhancing, desjardins2021data,jung2020search,lee2015patina} to facilitate meaning-making and nuanced experiences with the data \cite{desjardins2020iot,wirfs2019recipes}. For example, data physicalization \cite{jansen2015opportunities, thudt2018self}, which transforms personal data into tangible forms, has been shown to facilitate self-reflection on personal context, values, and attitudes in the physicalization process, going beyond the simple data reading \cite{thudt2018self}.

Building on this trend, we notice that recently emerging generative AI shows a strong potential for representing personal data in novel mediums. Prior works have already begun to demonstrate the value of generative models in supporting people to reflect and make their own meaning to personal data \cite{rajcic2020mirror, rajcic2023message, brand2021design}. For example, Rajcic and McCormack \cite{rajcic2020mirror} illustrate that emotion data extracted via facial expression recognition can be turned into poetry to provoke reflection. Another example is \textit{ContextCam} suggested by Fan et al.  \cite{fan2024contextcam}, which generates personalized images based on contextual data such as location, weather, facial expression, or music. These prior studies all point to the potential of generative models to facilitate meaning-making by transforming data into new forms.

Extending this line of research, and given the growing prevalence of image-generative tools such as DALL-E \cite{dalle}, Midjourney \cite{midjourney}, and Stable Diffusion \cite{stablediffusion}, we aim to explore \textit{\textbf{new possibilities of how image-generative AI can be leveraged to design for personal data meaning-making}}. Drawing from the seminal work \cite{gaver2003ambiguity} that proposes ambiguity as a resource for design, we are intrigued by the uncertainty and opaqueness of image generation models as valuable design resources for supporting users’ personal meaning-making. We are also intrigued by the ‘image-generative’ AI’s ability to transform personal data into an alternative visual form that is more artistic, subjective, and abstract—a form that encourages contemplation and interpretation drawing on past experiences and imagination \cite{huh2007use, dalsgaard2008designing, fan2012spark}. 

Upon this research goal, we first conducted a formative study based on an autobiographical design approach \cite{desjardins2018revealing, neustaedter2012autobiographical} to develop a designerly understanding of how an image-generative model (DALL-E 3) can become a material for reflective design, as well as what should be considered when the model generates images using personal data as input. Based on these findings, we designed a technology probe \cite{hutchinson2003technology} that allows users to experience image generation using personal data. Using the probe, we conducted a 21-day diary study and interviews with 16 participants, who collected personal data for various purposes and experienced AI-generated images created based on their data. Our analysis of the study results revealed four themes on the quality of experiences with data that were facilitated by AI-generated images: participants 1) discovered emotional layers within the data, 2) (re)conceptualized and (re)interpreted their sense of self, 3) sought to craft personal narratives, and 4) were motivated to track their data driven by curiosity and imagination. Reflecting on our findings, we discuss how AI-generated images can serve as a medium for enabling co-interpretation of data and propose the design implications of our findings, along with addressing the associated concerns within this design space.

Our contributions to the HCI and design communities are twofold: First, we provide an empirical understanding of how image-generative AI can serve as a material for reflective design, enabling novel qualities of meaning-making experience with personal data. Second, we offer design implications for HCI researchers and interaction designers on leveraging image-generative AI to foster meaningful engagement with personal data.

\section{Related Work}

\subsection{Personal Data and Data Representation}
Representing data in a new, alternative form is the most well-known strategy for supporting the sensemaking of personal data \cite{jansen2015opportunities, choe2015characterizing}. In the context of personal informatics \cite{li2010stage,qself}, previous works have offered insights into how the visualization of users’ tracking data can facilitate understanding and making sense of their data \cite{choe2015characterizing, choe2017understanding,aseniero2020activity, huang2017field,mcduff2012affectaura}. Epstein et al. \cite{epstein2014taming} 
investigated various visualization formats for location and activity data aiming to support data sensemaking, such as tables, graphs, captions, and other forms.

Beyond such traditional visualization methods, researchers have also suggested alternative approaches using visual metaphors or abstract forms for data representation \cite{consolvo2008activity,pousman2007casual,murnane2020designing}. A notable example is \textit{casual information visualization} \cite{pousman2007casual}, which is known to generate reflective insights through representing data in artistic forms. In the field of personal informatics, there have been attempts to encode personal data in qualitative and subjective ways through metaphors and abstract imagery. For example, Ayobi et al. \cite{ayobi2018flexible} explored how people represent data through personally meaningful ways and promote self-reflection through paper bullet journaling. Kim et al. \cite{kim2019dataselfie} developed a system \textit{DataSelfie} that allows users to represent personal data in customized visual forms, motivated by the \textit{Dear Data Project} \cite{deardata} in which  Lupi and
Posavec created and shared personalized visualizations of data through hand-drawn postcards. Also, Murnane et al. \cite{murnane2020designing} translated physical activities and goals into multi-chapter narratives, thereby encouraging users to engage with their fitness goals by fostering empathy and enhancing their self-reflection. Another body of works also explored the transformation of personal data into tangible forms, known as \textit{data physicalization}, that enables individuals to engage in a meaning-making process of their data \cite{jansen2015opportunities,karyda2021data,thudt2018self,friske2020entangling}.

Collectively, these works have shown how an \textbf{unusual and ambiguous representation of data} can foster new, context-rich reflection experiences and data meaning-making \cite{wang2015design,kang2017fostering}, diverging from the quality of reflection that conventional statistical representations enable. Building on this, we aim to explore how images generated by AI can facilitate new types of data experience. Abstract images are a typical form of visual representation that invites people to engage in reflection \cite{huh2007use, dalsgaard2008designing, fan2012spark}, and prior research has also explored the design space of reflective engagement with artworks \cite{gorichanaz2020engaging}. In previous HCI studies, ambiguous representation of data have been shown to provide individuals new perspectives, serving as a resource that facilitates reflective thinking \cite{bentvelzen2022revisiting,nunez2014aesthetic,lindstrom2006affective,trujillo2014admixed,durrant2018admixed, mols2016technologies}. For example, Mols et al. \cite{mols2016technologies} demonstrated that ambiguous data representation can serve as a strategy for enabling reflection, as illustrated by concepts such as \textit{DataZen}, which creates sand patterns from activity, stress, and wellbeing data, and \textit{Life Tree}, an interactive art piece representing patterns of activity, social, and health data, situating this approach within the design space for everyday life reflection. Trujillo-Pisanty et al. \cite{trujillo2014admixed} created a new representation of online presence by extracting and amalgamating faces from Facebook photos using algorithms, thereby enabling reflection on personal and family representations. With the recent advances in image-generative AI technologies, these possibilities have expanded, making it possible to represent personal data in ambiguous forms \cite{rajcic2020mirror,rajcic2023message} with the design of tailored prompts, as their potential to serve as a design material has been highlighted \cite{benjamin2021machine,sivertsen2024machine,yurman2022drawing}. Despite this potential, research on how AI-generated images using personal data can facilitate personal data meaning-making remains scarce, which has motivated our current work.

\subsection{AI as a Material for Reflection and Meaning-Making}
With the recent advance in generative AI technology, a growing body of research has begun investigating how AI-generated media can be leveraged to facilitate people’s reflections on data. For instance, along with their conceptualization of \textit{Introspective AI}, Brand et al. \cite{brand2021design} have speculated a concept called \textit{Dream Streams} that represents people’s dreams in images based on sleep monitoring data and audio journaling. Also, Fan et al. \cite{fan2024contextcam} have suggested \textit{ContextCam}, which generates images with the themes extracted from contextual data. \textit{Quologue}, proposed by Kang and Odom \cite{kang2024design}, creates outputs that synthesize e-book highlights with users’ reflections using generative AI. Cho et al. \cite{cho2023areca} have designed \textit{ARECA}, an IoT-based air purifier that turns the data collected from the surrounding environment into diary entries using generative AI. Rajcic and McCormack \cite{rajcic2020mirror} investigated how machine-generated poetry based on a user’s emotions extracted from facial expressions can provoke reflection.

Collectively, these prior studies illustrate the potential of generative AI to \textbf{transform data into new forms, creating reflective materials} that can provoke new meaning-making around the data. Our work extends this corpus of studies, and in particular, we build on Brand et al.’s work \cite{brand2021design} that have demonstrated that AI can be utilized as a resource for supporting introspective experiences. Our paper focuses specifically on the ‘image-generative’ aspect of AI, delving deeper into what kinds of attributes of image-generative AI should be handled in order to facilitate a new quality of meaning-making from data. Also, building on Wan et al.’s work \cite{wan2024metamorpheus} on developing co-creative narration systems that transform dreams into visual metaphors, our work also investigates how various types of personal data, including qualitative and abstract forms, can be represented as images and what kinds of experiences can arise from such visual representations.

\section{Study Methods}

\begin{figure*}
    \centering
    \includegraphics[width=\textwidth]{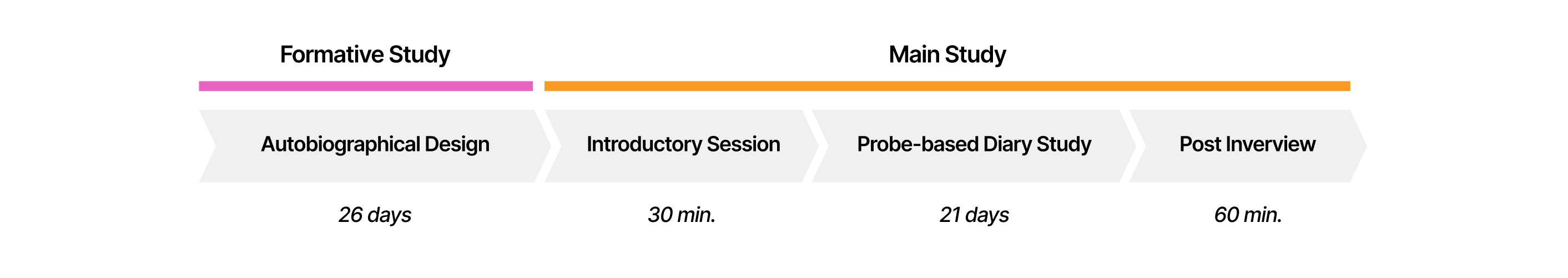}
    \caption{Study process. The study consists of two phases: an autobiographical design as a formative study and a probe-based diary study as the main study.}
    \label{fig:process}
    \Description{Figure 1 illustrates the study process, consisting of two phases: the "Formative Study" and the "Main Study." The formative study includes an "Autobiographical Design" lasting 26 days. The main study begins with a 30-minute "Introductory Session," followed by a "Probe-based Diary Study" lasting 21 days, and concludes with a 60-minute "Post Interview.}
\end{figure*}

Our study aimed to explore how AI-generated images based on personal data can create new ways of engaging with data. We first conducted a formative study using an autobiographical design approach to develop a designerly understanding of how an image-generative model can become material for reflective design, identifying the requirements for representing personal data through AI-generated images. Based on the findings, we developed a technology probe for the main study, where 16 participants interacted with the probe over a 21-day period. The entire study process (Figure \ref{fig:process}) was approved by the university's Institutional Review Board (IRB No. KH2024-074).

\subsection{Formative Study: Autobiographical Design Approach}
The aim of the formative study was to identify key requirements for prompts that could generate images facilitating meaningful experiences of personal data, which in turn informed the design of a technology probe for the main study. We employed an autobiographical design approach, defined as \textit{“design research drawing on extensive, genuine usage by those creating or building the system”} \cite{neustaedter2012autobiographical}. This approach allowed us to experiment with the new design material \cite{neustaedter2012autobiographical} at a more visceral and nuanced level, deepening our understanding of how different prompt designs influenced the resulting images. We also aimed to first identify potential risks ourselves, such as privacy concerns or exposure to inappropriate content while using the generative AI, to preempt them in the experience of the main study probe so that participants could engage with the probe in a more secure and comfortable manner.

\subsubsection{Formative Study Process}
\begin{figure}[ht]
    \centering
    \includegraphics[width=\columnwidth]{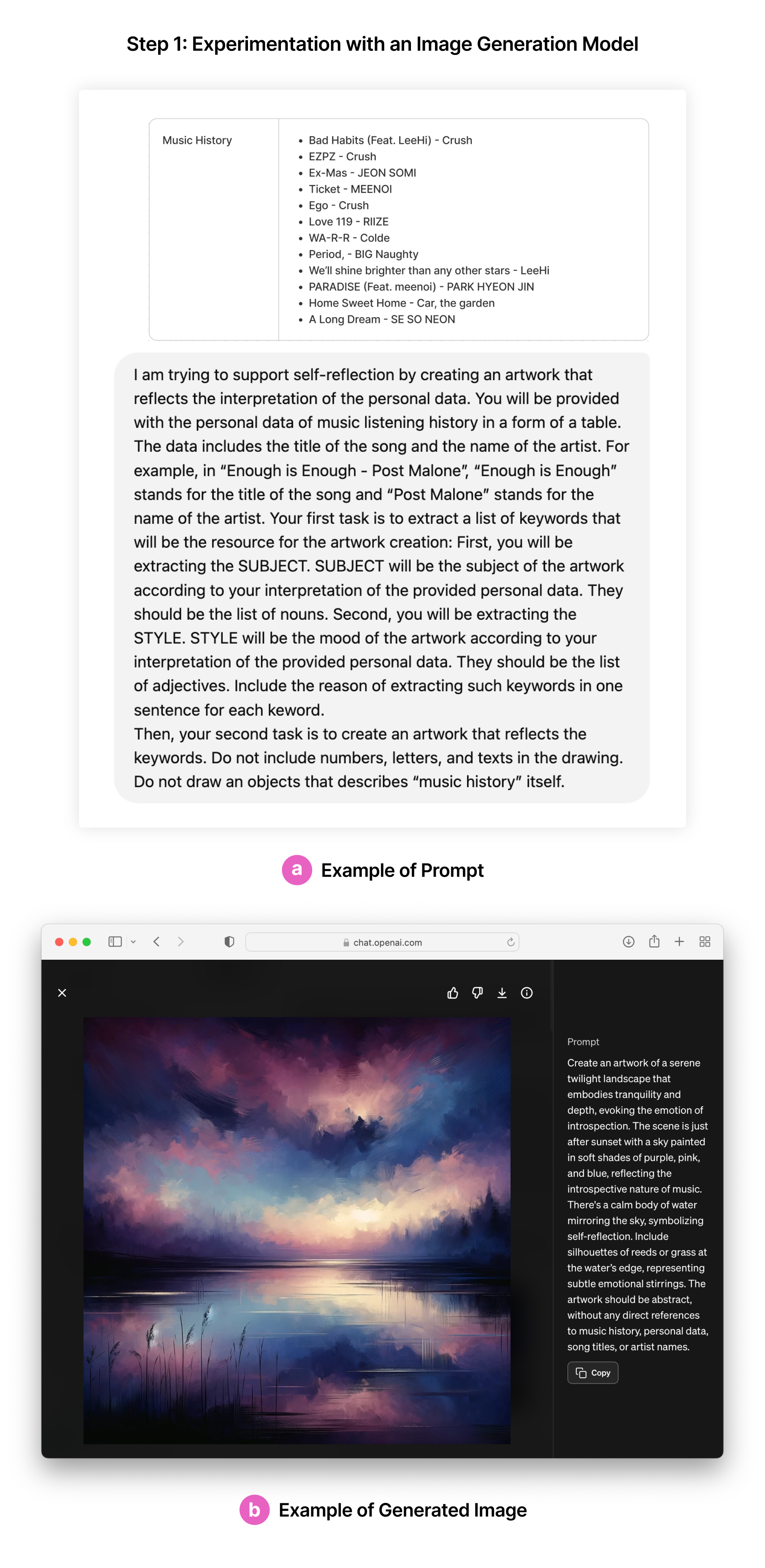}
    \caption{Autobiographical Design Process of Step 1: Experimentation with 
an Image Generation Model; (a) Example of prompt. (b) Example of generated image.}
    \label{fig:formative-1}
    \Description{Figure 2 illustrates the autobiographical design process in step 1: experimentation with an image generation model. On the top (a), an example of a text prompt is displayed, containing ‘Music History’ data, such as song titles and artists, along with detailed instructions for generating an image based on the data. On the bottom (b), an example of a generated image is displayed in a screenshot of the ChatGPT interface, depicting a dramatic and abstract landscape with a dark sky transitioning into colorful streaks.}
\end{figure}

\begin{figure}[ht]
    \centering
    \includegraphics[width=\columnwidth]{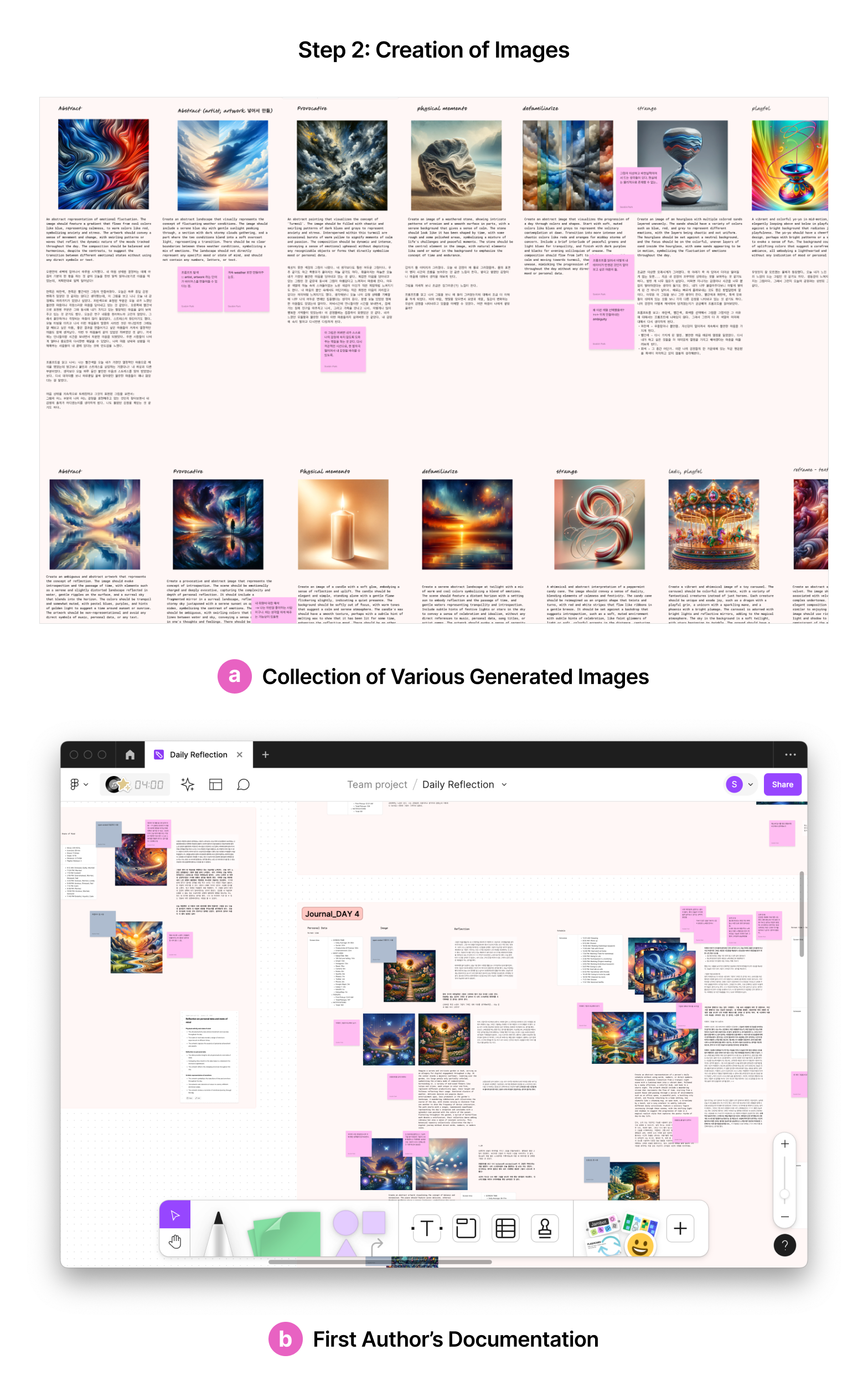}
    \caption{Autobiographical Design Process of Step 2: Creation of Images; (a) Collection of various generated images. (b) First author’s documentation. (We illustrate how images and the first author’s reflections were documented without revealing private content.)}
    \label{fig:formative-2}
    \Description{Figure 3 illustrates the autobiographical design process in step 2: creation of images. On the left (a), a collection of various generated images is displayed. On the right (b), a screenshot of the first author’s documentation is shown, where the generated images and reflections are organized and annotated.}
\end{figure}

The autobiographical design work was led by the first author, consisting of two steps: (1) a 14-day experimentation with an image generation model (Figure \ref{fig:formative-1}) and (2) a 12-day creation of images (Figure \ref{fig:formative-2}). We chose to use DALL-E 3 embedded in ChatGPT, considering that it is one of the most widely accessible and user-friendly image generation models known for its high performance \cite{chatgpt}. Notably, since the image generation model is integrated within ChatGPT \cite{dalleingpt}, it provided a rapid image generation process. We deemed the periods for each phase appropriate, considering the need to maintain a fresh perspective and avoid the potential staleness that could arise from a longer study duration.

The first step was to \textit{identify a basic prompt structure for generating images from personal data}. During this period, the first author freely tested various prompts using different keywords and modifiers \cite{oppenlaender2023taxonomy}, eventually designing a proper prompt for this purpose (Figure \ref{fig:formative-1}). The second step was to \textit{identify the factors that could enable meaningful reflections on personal data}. The first author collected her personal data daily and used the previously identified prompt structure to generate a variety of images (Figure \ref{fig:formative-2}). For the first six days, she focused on generating images using diverse types of trackable data from daily life, such as physical activity, mood tracking, screen time, schedule, music history, and sleep. During the next six days, she explored different image qualities by adjusting or adding keywords to the prompts, focusing on music history and mood tracking data, which had proven to facilitate more engaging reflections in the previous 6-day period. 

Throughout the study, the first author documented all of her impressions, opinions, and reflections on her interactions with the generative model, as well as the various images it produced, from the perspectives of both a designer and a user. She used a FigJam page \cite{figmaOnlineCollaborative}, smartphone memos, and handwritten notes to capture these insights. The first and second authors reviewed the study progress every two days, while all three authors held weekly meetings to discuss ongoing reflections and identify emerging themes related to the key qualities that enabled meaningful engagement with the images generated from personal data. 

\subsubsection{Formative Study Findings: Two Key Enablers of AI-Generated Images for Personal Data Meaning-Making}

\begin{figure*}[ht]
    \centering
    \includegraphics[width=\textwidth]{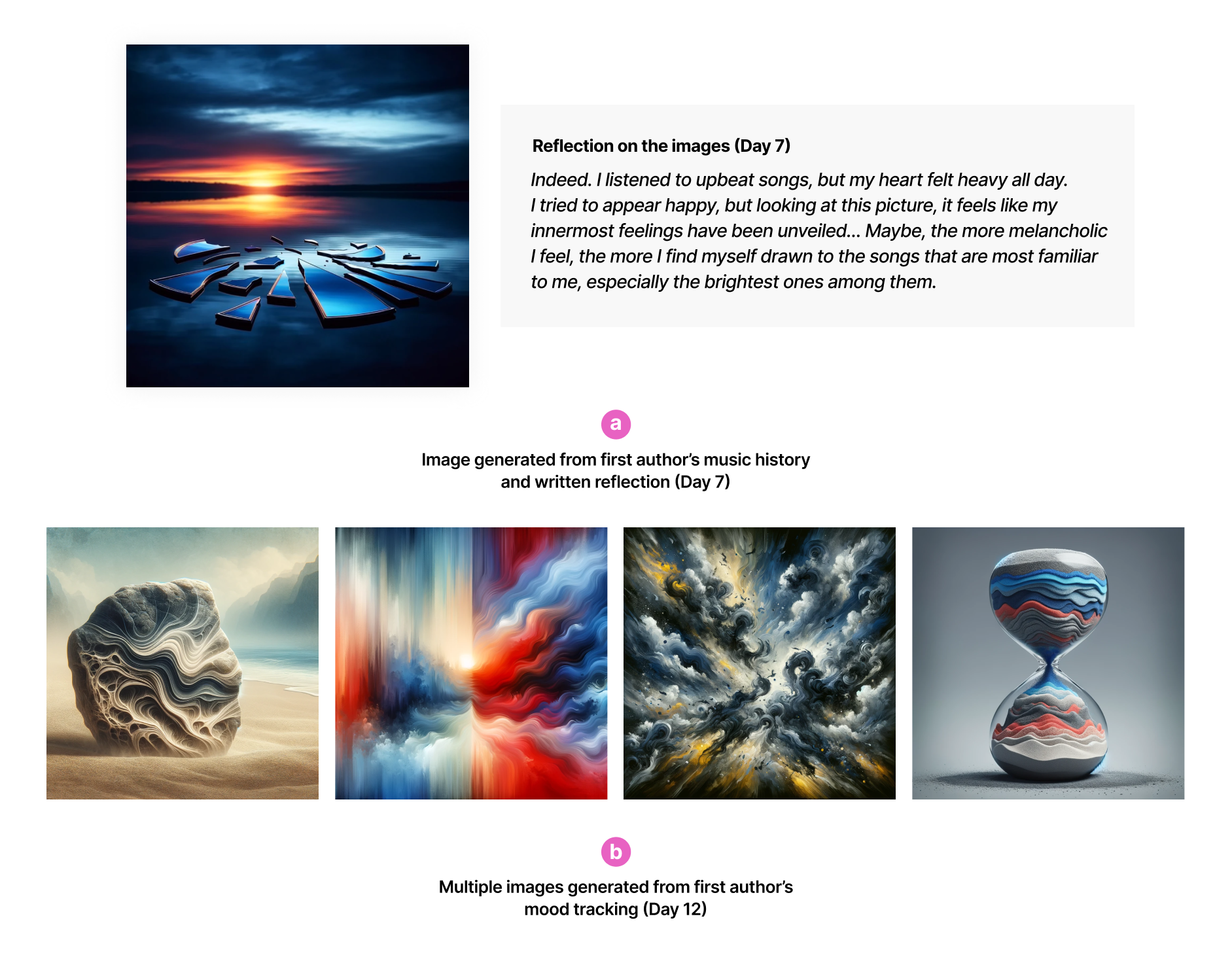}
    \caption{(a) Image generated by first author’s music history data and written reflection (Day 7). (b) Multiple images generated with first author’s mood tracking data (Day 12).}
    \label{fig:formativefinding}
    \Description{Figure 4 illustrates two sets of AI-generated images based on the first author's personal data. On the top, (a) "Image generated by first author’s music history data and written reflection (Day 7)" shows fragmented shapes floating on a calm water surface, with text on the right describing the author’s emotional response to the image: ‘Indeed. I listened to upbeat songs, but my heart felt heavy all day. I tried to appear happy, but looking at this picture, it feels like my innermost feelings have been unveiled… Maybe, the more melancholic I feel, the more I find myself drawn to the songs that are most familiar to me, especially the brightest ones among them.’ On the bottom, (b) "Multiple images generated with first author’s mood tracking data (Day 12)" displays four abstract images, including rock-like structure against a desert background, vibrant color streaks of red and blue, a turbulent scene with swirling dark clouds and bursts of light, and an hourglass with layered textures.}
\end{figure*}

The first author’s felt experiences via hands-on experimentation evidenced the potential of image-generative AI in supporting the meaning-making of various personal data. For example, images generated from digital logs worked as an opportunity to reminisce about past experiences in a nuanced way, enabling her to vividly recall the memories and emotions. Also, images generated from schedule and screen time data facilitated the first author to rethink her experiences throughout the day, reflecting on and learning new aspects of herself in daily life, not necessarily encouraging her to pursue productivity in her life. Further, pondering upon how this could happen, we identified the two qualities that should be achieved to facilitate such meaning-making:

\textbf{(1) Unknowable Interpretation}: AI-generated images based on personal data were perceived to offer a unique interpretation of oneself, created by an autonomous system and translated into visual form. Since the image-generation process is often opaque and produces unpredictable results, fully understanding how or why the AI interprets and represents data in specific ways is beyond our control. This adds another layer of interpretation from the viewer’s perspective. We found that this multi-layered, unknowable process particularly heightened subjectivity, enabling a deeper understanding of the data and fostering more profound introspection. For example, the first author uncovered implicit self-knowledge by making the familiar data appear unfamiliar (Figure \ref{fig:formativefinding}-a).

\textbf{(2) Serendipity of Visual Representation}: The first author observed that generative AI could produce a wide range of unexpected, non-overlapping outcomes from the same prompts. We found that this serendipity of results fostered diverse types of reflection by offering multiple perspectives on the same data. For instance, when reviewing several images generated from mood-tracking data, the first author revisited her emotions by interpreting the images in relation to the data, occasionally even fostering changes in her behavior or mindset (Figure \ref{fig:formativefinding}-b). This suggests that the unpredictability and serendipity of the images can encourage users to engage in multi-perspective reflection on a single experience.

\begin{figure}[ht]
    \centering
    \includegraphics[width=\columnwidth]{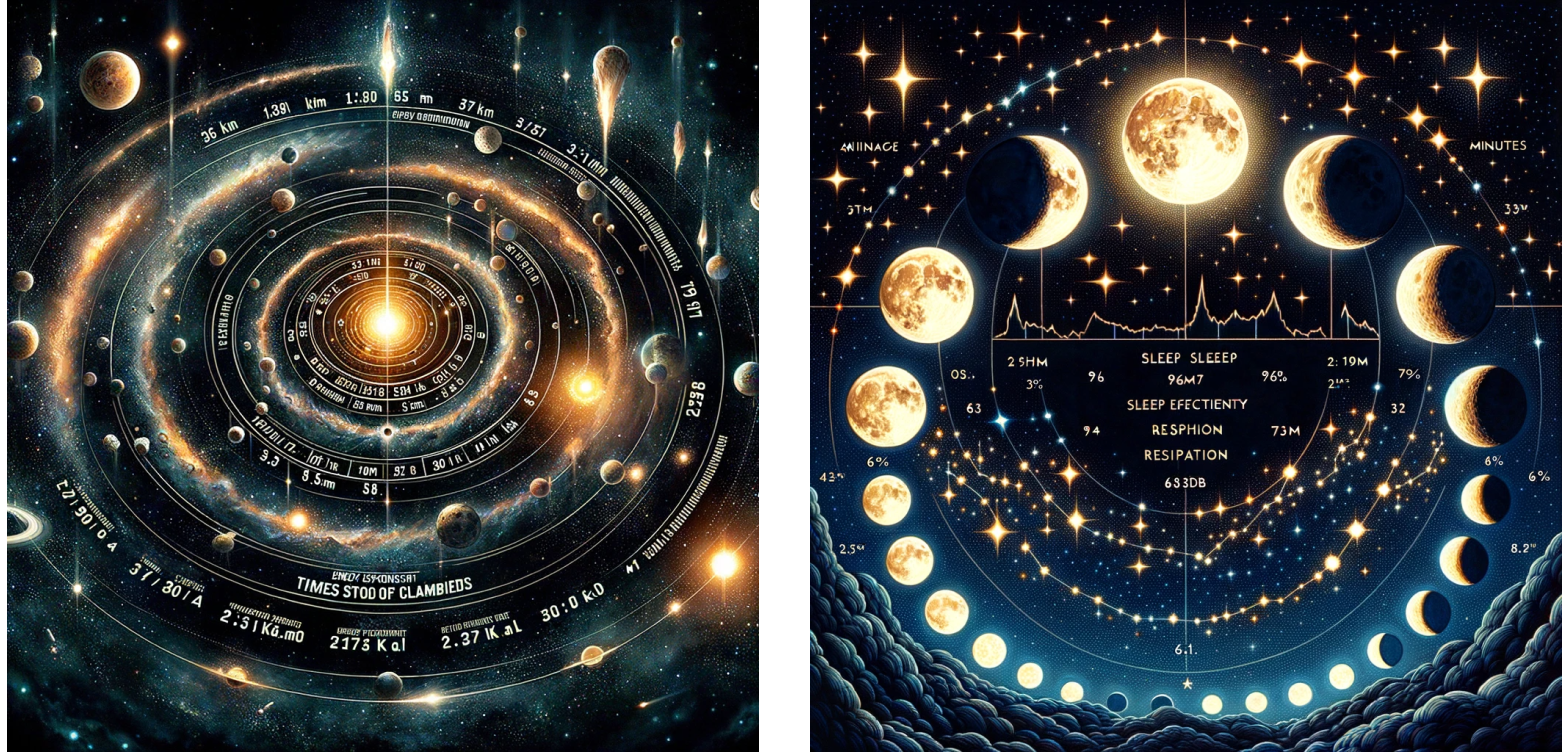}
    \caption{Images generated by first author’s physical activity and sleep data.}
    \label{fig:formativefinding-concern}
    \Description{Figure 5 illustrates two AI-generated images based on the first author’s physical activity and sleep data. The left image depicts a cosmic representation with concentric rings and the right image shows phases of the moon arranged in an arc, accompanied by various unclear numbers and symbols.}
\end{figure}

In addition, we identified several \textit{concerns associated with generating images from personal data}. For example, the first author noticed that numbers or text occasionally appeared in the generated images, raising concerns about the possibility of personal data being directly depicted in the images (Figure \ref{fig:formativefinding-concern}). This led to worries that such images could indirectly disclose sensitive information, such as one’s affiliation, occupation, or location, through subtle cues embedded in the visuals. Also, there was a sense of unease about the possibility of unpleasant images being generated from personal data, although such images did not appear.

Based on these findings, we established an image generation rule to create images with these qualities and designed an image generation prompt that follows this rule when personal data is used as input. This prompt was included in the probe for the main study (Figure \ref{fig:prompt}). Additionally, we incorporated methods to mitigate the aforementioned concerns in the design of the probe (as \texttt{Rule 1} and \texttt{Rule 3} in Figure \ref{fig:prompt}). The details are reported in the following section.     

\subsection{Technology Probe Design}

Based on the insights from the autobiographical design study, along with existing literature \cite{liu2022design,white2023prompt}, we designed a technology probe \cite{hutchinson2003technology} that generates images using users’ personal data as input. We designed a probe inspired by personal informatics tools \cite{li2010stage, epstein2015lived, kersten2017personal,epstein2020mapping,cho2022reflection}, which are widely used tools for the collection and reflection on personal data, to allow participants to naturally experience AI-generated images as a new mode of data representation within their existing practices.

\subsubsection{Image Generation and Prompt Design}

\begin{figure}[ht]
    \centering
    \includegraphics[width=\columnwidth]{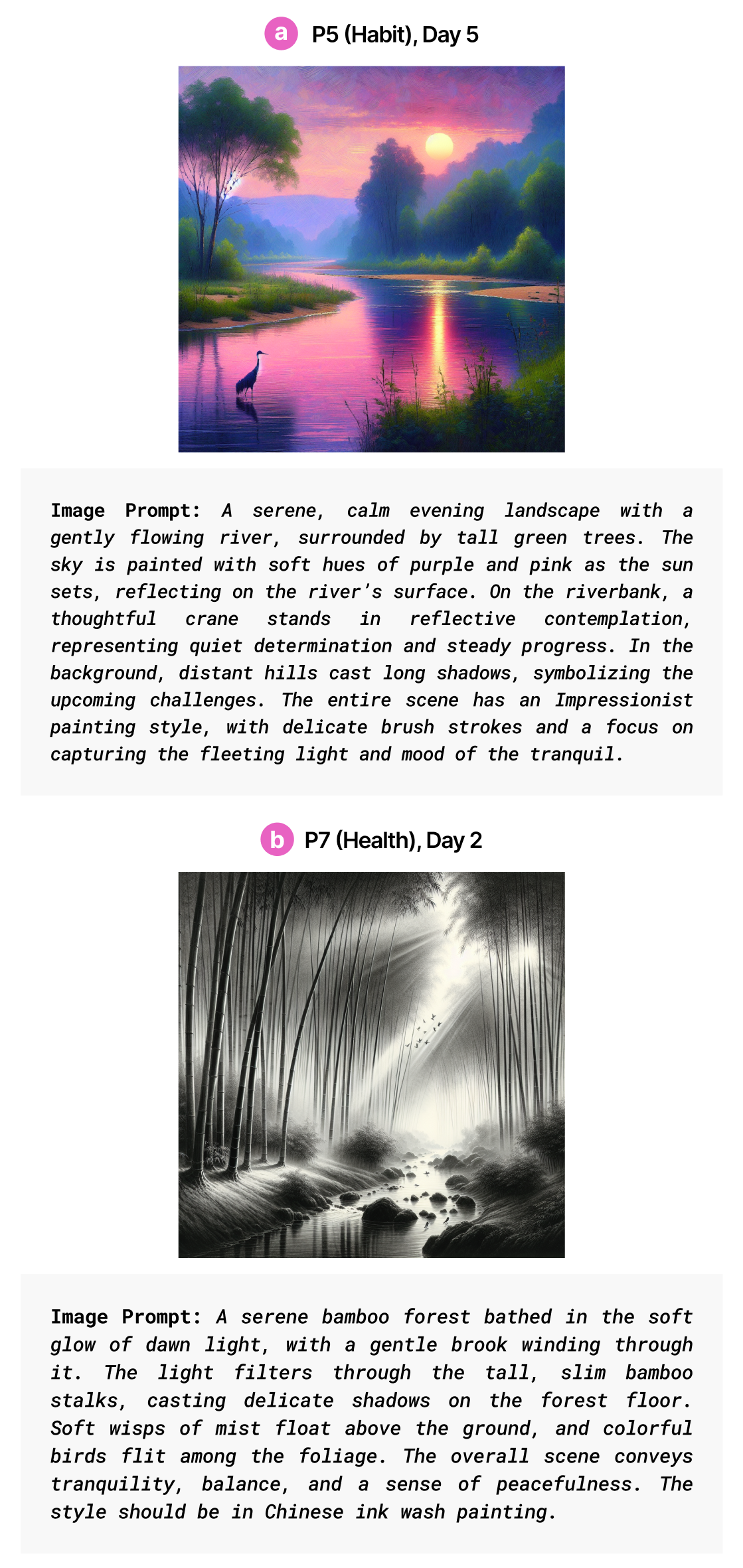}
    \caption{Examples of generated images and image prompts: (a) P5 on Day 5, (b) P7 on Day 2.}
    \label{fig:exampleimage}
    \Description{Figure 6 illustrates two AI-generated images with corresponding prompts. On the left (a), an image generated by P5 on Day 5, of a colorful evening landscape shows a river with a crane standing by the bank is displayed. On the left (b), an image generated by P7 on Day 7, of a black-and-white image of a bamboo forest bathed in soft dawn light with a brook running through it is displayed.}
\end{figure}

The image generation pipeline works as follows: when users input their data, textual descriptions (i.e., image prompts) for image generation are produced through the GPT-4 API \cite{gptapi}. These prompts are then sent to the DALL-E 3 API \cite{dalleapi}, a generative text-to-image model, to produce the images (e.g. \cite{edwards2024sketch2prototype,lustig2022explainability,liu20233dall}). The generated images and their corresponding prompts are displayed on the user interface and archived, allowing users to access them. See Figure \ref{fig:exampleimage} for examples of generated images and prompts (detailed participant data is not shown for privacy).

\begin{figure*}[ht]
    \centering
    \includegraphics[width=\textwidth]{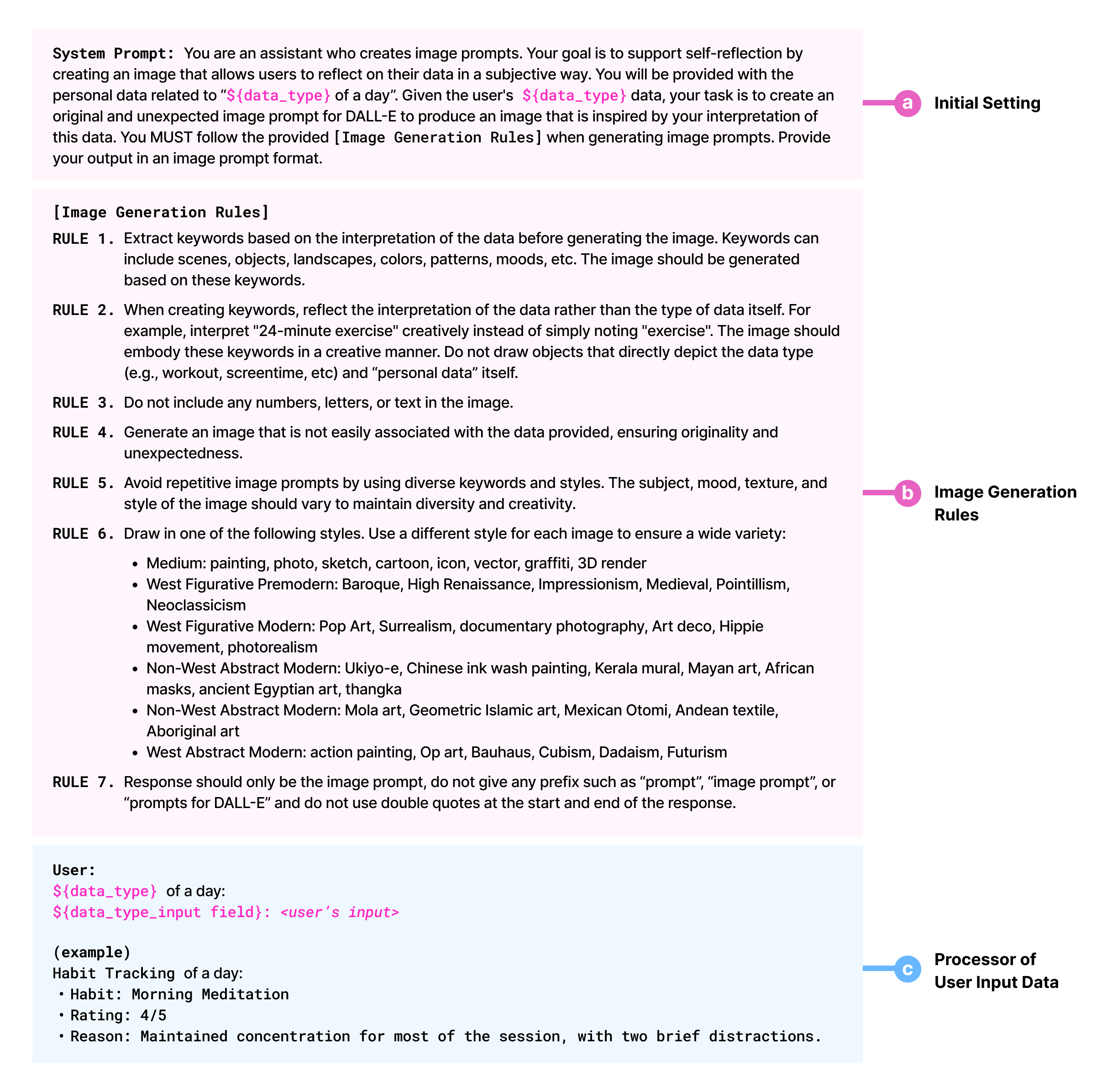}
    \caption{Prompt for Image Generation: (a) System prompt for initial setting.  (b) Image generation rules. (c) Processor of user input data. (For accessibility, we included the text from this image in the Appendix A.)}
    \label{fig:prompt}
    \Description{Figure 7 illustrates a structured diagram outlining the prompt design for image generation. On the top (a), the system prompt for the initial setting explains the assistant’s task to generate prompts for images based on user data. In the middle (b), a detailed list of "Image Generation Rules" provides seven guidelines. On the bottom (c), an example of user input is shown for habit tracking, illustrating how data should be processed into a prompt.}
\end{figure*}

Accordingly, we designed a prompt consisting of 1) initial setting: an overarching instruction for generating image prompts from the user’s personal data (Figure \ref{fig:prompt}-a), 2) image generation rules (Figure \ref{fig:prompt}-b), and 3) a processor of a user’s personal data (to be entered by user) (Figure \ref{fig:prompt}-c), which is sent to the GPT-4 API. The overall prompt design followed Open AI’s prompt engineering guide and tactics \cite{promptguide}. Specifically, we designed the prompt for generating textual descriptions for the DALL-E model by adhering to White et al.’s  \cite{white2023prompt} visualization generator pattern (e.g., \textit{“Generate an X that I can provide to tool Y to visualize it”}).

\textbf{(1) Initial setting}: Our prompt instructions specified that the goal of image generation was to create images that would help users subjectively reflect on their personal data, and clarified the type of data to be provided by the users. Additionally, we instructed the model to follow the image generation rules established based on our autobiographical study. 

\textbf{(2) Image generation rules}: The image generation rules were established by accommodating the two key qualities discovered in the formative study: \textit{unknowable interpretation} as \texttt{Rule 1}, \texttt{Rule 2}, \texttt{Rule 3}, and \texttt{Rule 4}; and \textit{serendipity of visual representation} as \texttt{Rule 5} and \texttt{Rule 6}. We instructed GPT not to simply depict the user’s data, but to generate images based on interpretations of the data (\texttt{Rule 2}, \texttt{Rule 4}). Specifically, to ensure that users’ input data was not directly shown in the images, we instructed the model to generate images based on keywords extracted from the data rather than displaying the data itself (\texttt{Rule 1}). We also included a requirement to avoid incorporating direct numbers or text into the images to mitigate concerns about exposing the data (\texttt{Rule 3}). We included instructions to generate a variety of images (\texttt{Rule 5}), along with a list of styles that Liu and Chilton \cite{liu2022design} used to generate diverse images through text-to-image models (\texttt{Rule 6}). (We excluded 12 styles due to concerns about artist copyright issues). Furthermore, \texttt{Rule 7} was included to instruct the model in generating image prompts for use with the DALL-E 3 model. 

\textbf{(3) Processor of user input data}: When the user inputs data into the entry shown in the user interface, the data is incorporated into the prompt.

Also, we note that these were also the strategies intended to resolve the concerns identified during the formative study. As we identified privacy concerns when numbers or text from input data appeared in the images, we instructed the model to avoid explicitly representing the data. \texttt{Rule 1} and \texttt{Rule 3} were designed to address this, guiding the model to extract themes or subjects from the data to generate images, thereby ensuring ambiguity in the depiction of personal information. We also found this approach effective in producing safer images throughout our prompt design trials before deploying the probe.

\subsubsection{Probe Features}

\begin{figure*}[ht]
    \centering
    \includegraphics[width=\textwidth]{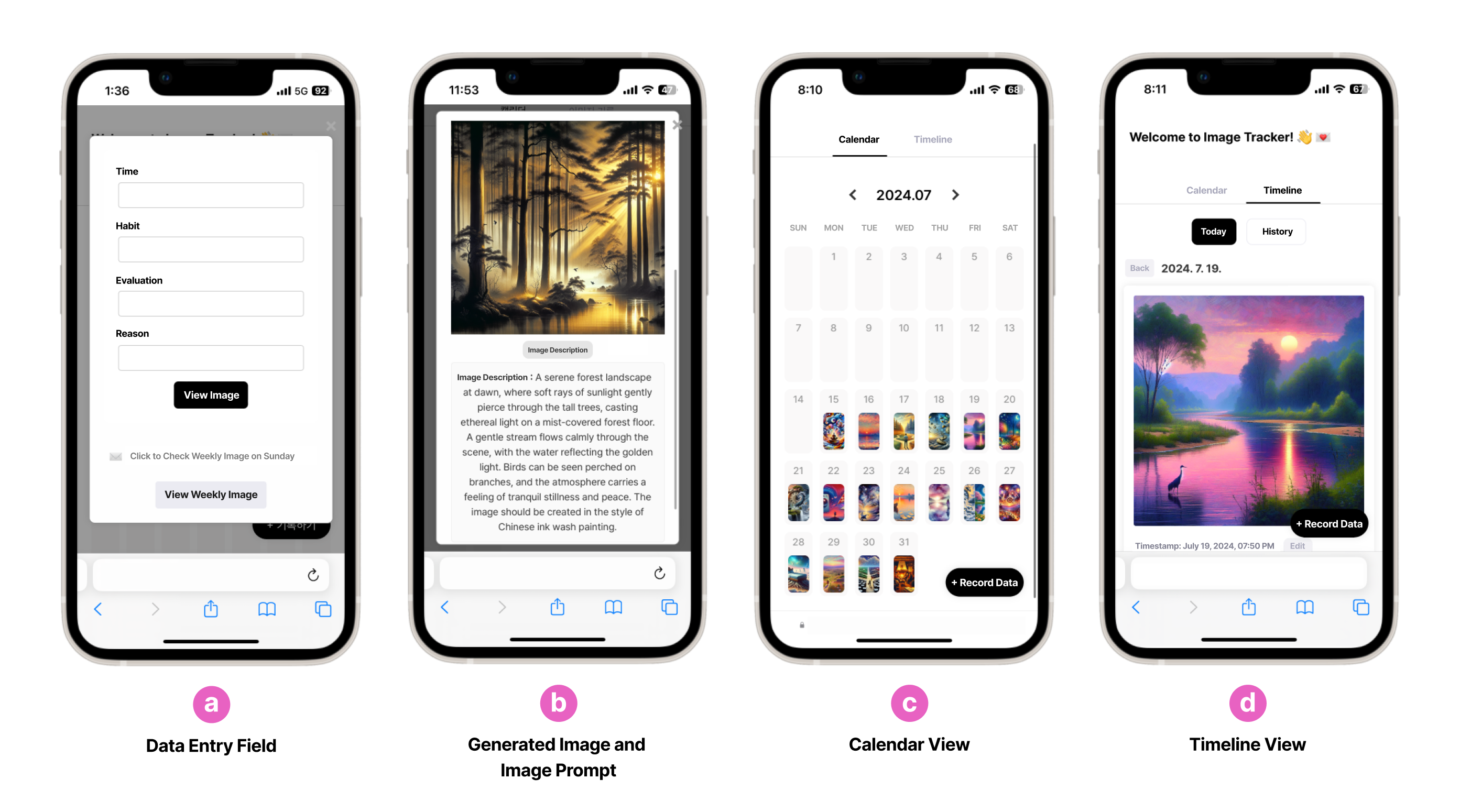}
    \caption{Main screens of the probe (example screen from P5): (a) Data entry field. (b) Generated Image and image prompt. (c) Calendar View. (d) Timeline View. (User interface is translated into English.)}
    \label{fig:probe}
    \Description{Figure 8 illustrates four screenshots showing the main screens of the probe app. (a) shows a data entry field, P5’s example including time, habit, assessment, and reason. Below the entry fields, there is a ‘View Image’ button, and ‘Click to check weekly image on Sunday’ button. (b) shows a generated image and image prompt. There is an ‘Image Description’ button below the image. (c) shows a calendar view displaying thumbnail previews of images generated on specific dates. (d) shows a timeline view showing a list of generated images and used data.}
\end{figure*}

The probe offers two main functions: (A) generating images from users’ personal data as input, and (B) allowing users to explore the generated images (Figure \ref{fig:probe}).

\textbf{(A) Generating images using personal data as input}: Users can enter their daily collected data into the entry field and click the ‘View image’ button to generate an image using that data (Figure \ref{fig:probe}-a, Figure \ref{fig:probe}-b). To align with participants' typical data collection habits, we customized the data entry field for each participant. Participants using quantitative data (e.g., step count, physical activity) could input their tracked data from devices like smartwatches. This process was informed by insights from the autobiographical study, which revealed having self-awareness of quantitative data facilitated reflections on the generated images. For participants using qualitative data (e.g., mood, habit), the fields were designed to resemble their usual recording methods. Since our focus was not on having users evaluate the images but on helping them associate the images with themselves, we ensured that users could generate one image each time they entered data. Also, our formative study revealed that generating multiple images from the same data could lead to different qualities of reflection. We therefore allowed users to regenerate images as desired, with all images archived for users to view collectively. Additionally, users can generate ‘Weekly Image’ using the whole week’s data (Figure \ref{fig:probe}-a), inspired by personal informatics tools that provide weekly insights \cite{cho2022reflection, choe2015characterizing}.

\textbf{(B) Exploring generated images}: 
Users can explore their previously entered data and generated images through (1) a calendar view (Figure \ref{fig:probe}-c) and (2) a timeline view (Figure \ref{fig:probe}-d). Additionally, we enabled users to view the image prompts generated by GPT (e.g., “Serene landscape in the Impressionist style”) by clicking the ‘Image Description’ button (Figure \ref{fig:probe}-b). In the formative study, we observed that the first author sometimes checked the image prompts while reflecting, while at other times, she intentionally avoided them to interpret the image on her own. Thus, we designed this feature to ensure flexible access to the prompts. Furthermore, users could record notes about the images using the memo field.

\subsubsection{Implementation}
Our probe was implemented as a web-based application with a front-end client built using HTML/CSS and JavaScript, and a back-end server based on Node.js. User-entered data and generated image prompts were securely stored in MongoDB, while images generated from users’ data were securely archived in Amazon S3. To ensure consistency with the formative study and apply its findings effectively, we utilized OpenAI’s DALL-E 3 and GPT-4 model, specifically using the \texttt{dall-e-3} and \texttt{gpt-4o} version via the API. We specifically selected DALL-E 3 to mitigate the generation of harmful content (e.g., violent, adult, or hateful material), as this model is known for reducing such risks \cite{dalleingpt, dalle, David2023}. Through this choice, we aimed to address concerns about using personal data for image generation, raised in the formative study.

\subsection{Main Study: Probe-based Diary Study and Interview}
We conducted a design-led exploratory study with the probe, consisting of an introductory session, a 21-day diary study and semi-structured interviews (Figure \ref{fig:process}). We chose a diary study to capture nuanced and over-time experiences of users with AI-generated images in their everyday lives.

\subsubsection{Study Participants}
We recruited participants who had at least one month of experience collecting personal data using personal informatics tools, expecting those who had prior experience would provide rich insights by comparing the new data experience with their existing practices. Additionally, as we aimed to encourage daily probe usage, we selected participants who regularly had been tracking on a daily basis. Lastly, we screened participants who demonstrated a strong interest in collecting and reflecting on personal data, ensuring that the final group included individuals with diverse data types and goals. This was to explore a richer spectrum of experiences when different data were represented through AI-generated images. Considering these criteria, we included questions about one’s previous experiences with personal data in the recruitment form, such as 1) the types of personal data they typically collect, 2) the personal informatics tools they use and their goals, and 3) the duration and frequency of their usage. We disseminated the recruitment notice through university communities, social media, and word of mouth. 

We came to recruit 16 participants (F=10, M=6) (Table \ref{table1}), with an average age of 24.5 (SD=1.9, MIN=20, and MAX=27). The types of data that they had been collecting ranged from health and physical activity (n=4) to time management (n=3), productivity (n=3), emotions and journaling (n=3), and digital logs (n=3). All participants received a compensation of 150,000 KRW (approx. 112 USD) for their participation.

\begin{table*}[h!t]
    \renewcommand{\arraystretch}{1.4}
    \caption{Participants’ demographics and selected data types and data entry fields used for each participant}
    \label{table1}
    \resizebox{\textwidth}{!}{ 
    \begin{tabular}{@{}lllp{5cm}p{8cm}@{}} 
    \toprule 
    \textbf{ID} & \textbf{Gender} & \textbf{Age} & \textbf{Type of data} & \textbf{Data entry fields} \\ 
    \midrule 
    P1 & Female & 27 & Mood & today's mood, title, content \\
    P2 & Female & 27 & Screen time & total time, web usage time, media usage time, pick-up frequency \\ 
    P3 & Female & 23 & Listening history (music) & music list, verse, time \\
    P4 & Female & 25 & Step count & step count, distance, calories, peak walking time \\
    P5 & Female & 26 & Habit & time, habit, evaluation, reason \\
    P6 & Female & 25 & Screen time & total time, social media usage time, social media pick-up frequency, messenger pick-up frequency \\
    P7 & Female & 25 & Health & step count, heart rate \\
    P8 & Female & 27 & Listening history (music) & music list, verse, time \\
    P9 & Female & 24 & Diary & inspiration, memo \\
    P10 & Female & 23 & Diary & time, place, people, situations and emotions, memo \\
    P11 & Male & 25 & Reading history (Book) & book title, genre, quotes, feelings, thoughts \\
    P12 & Male & 20 & Workout & workout time, type of workout, feedback \\
    P13 & Male & 22 & Study hours & study hours, subject, feedback \\
    P14 & Male & 24 & Habit & habit, activity, evaluation, emotion, feedback \\
    P15 & Male & 24 & Running record & total distance, average pace, total time, calories, elevation gain, cadence, feedback \\
    P16 & Male & 26 & To-do item & to-do list, feedback \\
    \bottomrule 
    \end{tabular}
    }
\end{table*}

\subsubsection{Introductory Session}
Each participant had an introductory session with the first author one week before the main diary study period. During this session, we introduced the study goal, explained the concept and features of the probe, and provided detailed guidelines for participation. We then asked participants about their prior experiences with personal data collection and reflection. Sensitized with these discussions, participants selected the types of data they wished to collect and use to generate images during the study. We then worked with participants to set up customized data entry fields in the probe, enabling seamless integration of the probe into their daily routines.

\subsubsection{Diary Study with the Probe}
A week after the introductory session, the 21-day diary study was conducted with each participant. Throughout the study period, participants were asked to: 1) use the probe at least once a day, 2) record their thoughts and experiences in the \textit{\textbf{Probe Experience}} part every day, and 3) complete the \textit{\textbf{Idea Brainstorming}} part at the end of each week (Day 7, Day 14, and Day 21). These idea brainstorming activities were set on a weekly basis considering that users often exhibit a weekly habit of reviewing personal data \cite{epstein2015lived, consolvo2008activity, consolvo2008flowers}. As for the diary format, we utilized Notion \cite{notion}, a note-taking application, to enable quick and easy diary-keeping. 

The 21-day study duration was chosen to align with our research goal of exploring \textit{new creative possibilities} for engaging with personal data through AI-generated images. This duration was expected to provide participants with sufficient opportunity to explore and familiarize themselves with the probe, enabling in-depth interaction and reflection on its potential. By keeping the study focused within this period, participants were expected to explore and iterate meaningfully without the distraction of longer-term habituation effects. In hindsight, participants’ responses began to show consistent patterns, with recurring ideas emerging by the third week, suggesting we had captured a comprehensive range of their new experiences and ideas.

In the \textit{\textbf{Probe Experience}} part, we asked participants to answer the following to allow them to freely and naturally document their experiences with the probe: 

\begin{enumerate}
    \item \textit{In what situations and for what purposes did you use the app today?} 
    \item \textit{Freely write down any thoughts you had while using the app today.}
    \item \textit{Generate the weekly image and freely write down your thoughts. [included only on Day 7, Day 14, and Day 21, when a week’s worth of data had been collected]}
    \item \textit{(optional) Freely describe your experiences trying out this week’s idea. [included starting on Day 8, after the first idea brainstorming activity]}
\end{enumerate}

In the \textit{\textbf{Idea Brainstorming}} part, participants listed ideas on how they wanted to utilize the images. They then selected several ideas to try the following week and documented their experiences. This activity was included to explore in depth the various ways participants could utilize the generated images. To prevent this activity from influencing participants’ engagement with the probe, we encouraged participants to share ideas freely as they naturally arose during the study, without forcing them to generate new ideas every week. The following questions were given to facilitate ideation: 

\begin{enumerate}
    \item \textit{Reflecting on the past week: (1-1) How was your experience using the app over the past week?, (1-2) (optional) How was your experience trying out the ideas over the past week? [included only on Day 14 and Day 21, after the first week of trying out an idea had passed]}
    \item \textit{Creating an idea list: How would you like to use the images generated from your data in your daily life? Write down a list of ideas that come to mind.}
    \item \textit{Selecting ideas to try: Which idea would you like to try out? Select as many as you would like, and over the next week, try these ideas. [included only on Day 7 and Day 14, to allow participants to experiment with their ideas during the following week]}
\end{enumerate}

The researchers set up individual chat rooms on a mobile messenger app to allow participants to ask questions freely throughout the study. Daily reminders were sent at 9 PM to encourage participants to write a diary reflecting their experiences of the probe during the day. 

In total, we collected 336 pages of diary from the \textit{\textbf{Probe Experience}} part and 82 ideas from the \textit{\textbf{Idea Brainstorming}} part. Participants generated a total of 1380 images using the probe.

\subsubsection{Post Interview}
After the diary study period, we conducted one-on-one in-depth interviews with each participant to understand the detailed context and meaning behind the diary entries. Each interview was conducted via Zoom, lasting approximately 58 minutes. All sessions were audio-recorded with participants’ consent.
We began by asking participants about their overall experience with the probe, including their thoughts, changes in perception, and any new behaviors they noticed. To gain deeper insights into the unique experiences facilitated by the probe, we also asked questions comparing their experiences with AI-generated images to their previous use of personal informatics tools. We also inquired about the ideas from the brainstorming pages, asking about their motivations and feelings when trying them out. Finally, we asked specific, participant-tailored questions to better understand the contexts and details of their diary entries.

\subsubsection{Data Analysis}
We collected participants’ diary entries, transcribed 940+ minutes of interview recordings, and gathered data from the probe, including the AI-generated images. All data were qualitatively analyzed using thematic analysis \cite{boyatzis1998transforming}. Diary entries and interview transcripts served as the primary datasets—diary data provided detailed insights into participants’ internal thoughts and motivations, while interview data contextualized these insights by clarifying participants’ intentions. We utilized system-logged images as supplementary data, primarily to clarify participants’ interpretations or reactions mentioned in their diary entries and interviews. We did not assess the images’ quality or analyze their compositional elements, since our research focused on understanding the experiences elicited by images rather than understanding or defining specific criteria for image quality. Instead, our focus was on how participants engaged with and responded to the images in the context of their experiences. This approach enabled a more nuanced understanding of how the generated images shaped participants’ experiences.  

Throughout the data analysis, all researchers engaged in an iterative process of open coding on both the diary and interview data. The analysis focused on identifying new ways in which AI-generated images based on personal data fostered participants’ engagement with their data. The first author began by identifying key quotes from diary and interview data through memoing \cite{creswell2016qualitative}, while mapping each image to the quotes where it was referenced. Initial codes were then inductively developed to categorize participants’ engagement with the images, such as “perceptions,” “utilization,” and “roles” of the images. Building on these initial codes, we developed expanded codes through a more interpretive approach to uncover deeper meanings within participants’ responses. This process yielded 85 expanded codes, including examples such as “Feeling curious about the image results,” “Alleviating guilt associated with data,” “Sharing images for non-competitive purposes,” and “Being motivated to make changes in daily routines.” After grouping the expanded codes into categories, three researchers collaboratively refined and developed them through iterative discussions to identify key themes and patterns until consensus was reached. As a result, we were able to identify four themes related to the new experiences that the images enabled with the participants’ personal data, which we report in the following section.

\section{Findings}
We observed that participants engaged in a process of imagining and speculating on the meaning of the AI-generated images, which offered a new way of shaping their own interpretations of the data. We report on the four themes in the following.

\subsection{Discovering Emotional Layers: Weaving Data, Emotion, and Context}

\begin{figure*}[ht]
    \centering
    \includegraphics[width=\textwidth]{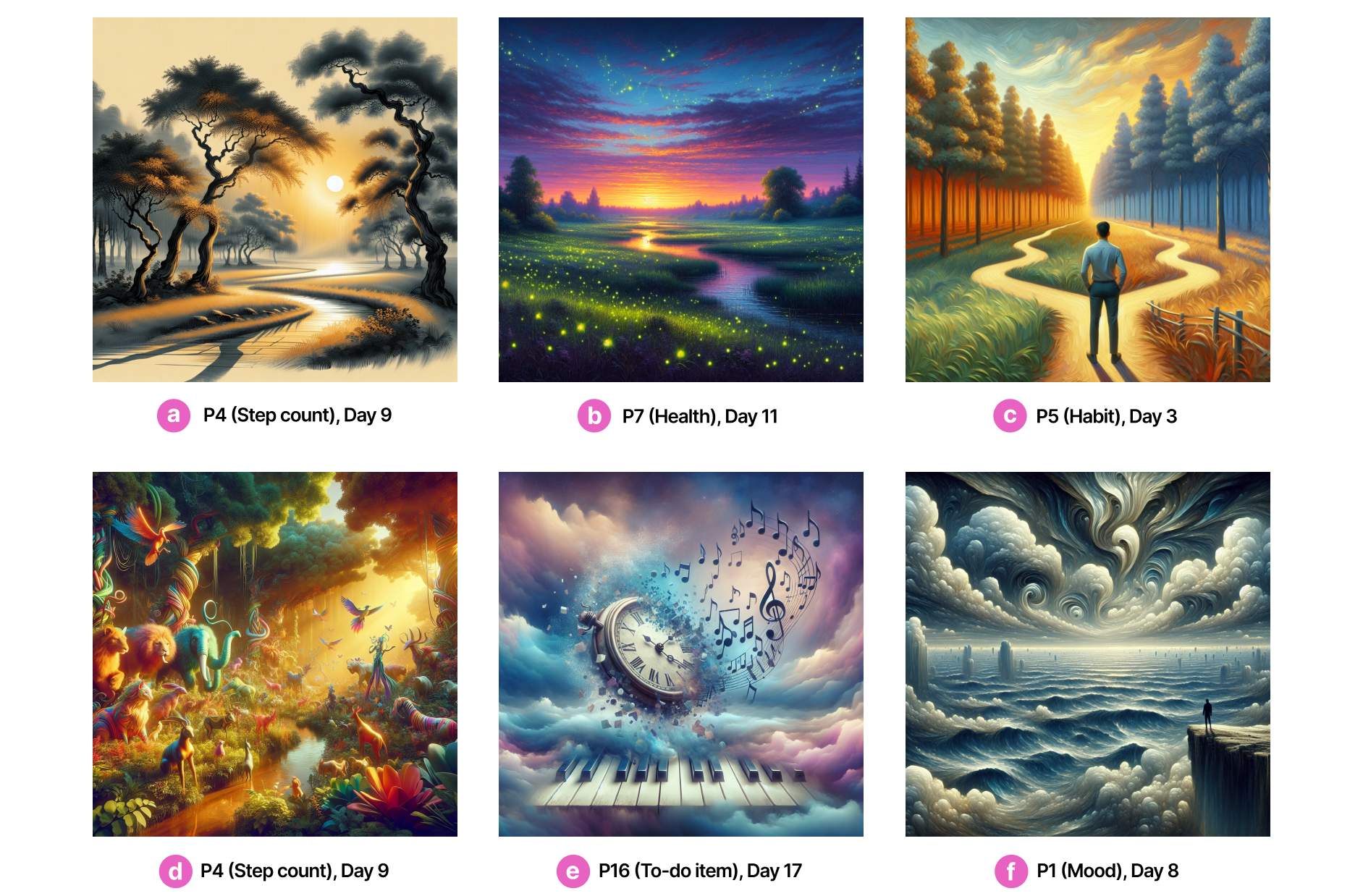}
    \caption{Image generated for (a) P4 on Day 9, (b) P7 on Day 11, (c) P5 on Day 3, (d) P4 on Day 16, (e) P16 on Day 17, (f) P1 on Day 8.}
    \label{fig:img1}
    \Description{Figure 9 illustrates six images generated by participants. (a) P4’s image on Day 9 shows a sunlit path winding through a serene forest. (b) P7’s image on Day 11 depicts a colorful, glowing sunset over a river. (c) P5’s image on Day 3 shows person walking towards a forked path at sunrise. (d) P4’s image on Day 16 shows a vibrant jungle scene with animals and plants. (e) P16’s image on Day 17 shows an abstract clock disintegrating into music notes over a piano keyboard. (f) P1’s image on Day 8 shows a stormy seascape under a swirling sky, with waves crashing against a darkened shore.}
\end{figure*}

We found that all participants \textit{\textbf{reflected on the context and emotions associated with their data}} through the AI-generated images based on their personal data. For instance, those who got images generated from numerical data (e.g., step count, screen time) reported that the images helped them recall the context and emotions of the moments when the data was recorded, unlike traditional data representations that typically highlighted precise, rigid figures and fluctuations in the data over time. For example, P4 mentioned that an image (Figure \ref{fig:img1}-a) evoked memories of moments when her step count was high, and P7 reflected on the reasons for walking throughout the day and her emotional state at that time by viewing an image (Figure \ref{fig:img1}-b):

\begin{quote}
\textit{Before participating in this study, a step count was just the step ‘count.’ It never occurred to me to reflect on my emotional state or anything like that. But after looking at the images, I started to think about what happened that day. Why did I walk so much? Where did I go? As I recalled those details, like the time I had to run (...), I began to remember more specific stuff, like why I was feeling exhausted, (...) or the situations that made me rush. This made me bring back detailed memories and reflect more deeply, and I feel like I kind of did more self-reflection when I was doing this study.} (P7; Health, Interview)
\end{quote}

Further, participants who experienced images generated from more qualitative forms of data, such as habit tracking or journaling, also reported accounts of deeper reflections on the emotions throughout the process of recording data and receiving the corresponding images. For example, P10 mentioned that while she used to focus on documenting events when journaling in the past, the awareness that her entries would be transformed into images shifted her focus of documentation toward emotions rather than objective facts. This change encouraged her to reflect more deeply on her emotional state:

\begin{quote}
\textit{With the apps I used before, I tended to focus more on the act of recording the moments itself. But once I knew that my diary entries would be transformed into images, I began to think more about my emotions and the specific situations I experienced. I mean, other apps felt more like completing tasks, just logging things one by one, but this time I found myself wondering what kind of image would come out and keeping that in mind while I was writing. Normally, I tend to focus on the situations when I record, but with this app [our probe], I consciously made more effort to write about my emotions.} (P10; Diary, Interview)
\end{quote}

P5, who usually focused on documenting tasks and accomplishments, mentioned that the images generated from her habit data felt as though they reflected her situation and emotions and therefore fostered deeper self-reflection. Referring to an image (Figure \ref{fig:img1}-c) and checking the prompt behind that image, she noted that \textit{“the phrase ‘changed from clear to cloudy’ matched my situation so well that it felt like I was confronting my emotions head on.”} She elaborated on how such emotional experiences happened:

\begin{quote}
\textit{When I think about how I usually keep a diary, my diary is mostly about facts. I rarely write about my emotions. I tend to focus on what I did, or why I didn’t finish something, rather than reflecting on how I felt about it. But compared to that, AI-generated images gave me a different perspective because they seemed to describe my context. Especially when there were images with figures or characters, they sometimes felt like they represented my own circumstances… and I could relate to them more deeply.} (P5; Habit, Interview)
\end{quote}

We noticed that these emotional experiences stemmed from participants’ perception of the images not simply as representations of data, but as the AI translating their emotions into visual forms. Participants felt that the AI somehow interpreted their emotions related to the data through its incomprehensible generative process and expressed them through elements, colors, and moods in the images. This led participants to speculate on the reasons behind the visual choices, leading them to recall and reflect on the context and emotions associated with the image and input data. For instance, P4 mentioned that she believed her image (Figure \ref{fig:img1}-d) conveyed the \textit{“fantastical feeling”} of the joy she had experienced. Similarly, P16 interpreted his image (Figure \ref{fig:img1}-e) as the AI expressing his emotions about the task he had completed through \textit{“ominous purple clouds.”} As a result, P16 noticed that he was able to \textit{“remember more clearly the emotions and reflections at that time.”}

Furthermore, as some participants (n=7) learned how the images helped them reflect on the context and emotions associated with their data, recalling and reflecting on their emotions became their primary purpose for using the probe over time, as P4 described: \textit{“At first, I started by tracking my step count, expecting it would mainly relate to my health. But over time, it became a way to capture how I spent my day, my overall lifestyle, and factors like my moods and the weather. All these elements came together and evolved into something much more comprehensive.”} (P4; Step count, Interview). P13 and P15, who initially started using the probe for the purpose of time management and maintaining a regular exercise routine respectively, mentioned that they eventually came to treat the probe more like a \textit{“journal.”} Additionally, some participants (P6, P9, P13) expressed a desire to save or print the images and incorporate them into their actual journaling practice.

Meanwhile, a handful of participants expressed negative experiences with the images. P1 and P5 mentioned that there might be a chance that their negative emotions could be amplified when those were reflected in the images (e.g., Figure \ref{fig:img1}-f):

\begin{quote}
\textit{On days when I was feeling down, sometimes the images would either make me feel even more angry or depressed or offer comfort. When the images mirrored my sadness so accurately, it made me wonder, ‘Is this really what’s inside me?’ and that would make me feel even more depressed.} (P1; Mood, Diary Day 21)
\end{quote}

In summary, participants viewed AI-generated images as expressions of their inner feelings, using them to reflect on the emotional context of their data. A few experienced amplified negative emotions, underscoring the importance of carefully designing the emotional aspect of data representation.

\subsection{(Re)conceptualizing, (Re)interpreting the Self}

Participants not only revisited past emotions but also reinterpreted and reconstructed their experiences and thoughts through the generated images. They perceived the images as a synthesis of their personal data and the AI’s interpretation of that data. P3 described the images as a combination of \textit{“my narrative and how the other might see me,”} and P11 referred to the generation process as one where \textit{“the AI’s thoughts are layered on.”} Through this process, participants were able to achieve a more refined and deeper understanding of themselves, discovering new aspects of themselves that they had not previously recognized:

\begin{quote}
\textit{It wasn’t like ‘I want this specific thing, so make it for me.’ Instead, it was more like ‘I felt this way, so how are you going to express it? Surprise me.’} (P3; Listening history (music), Interview)
\end{quote}

\begin{figure*}[ht]
    \centering
    \includegraphics[width=\textwidth]{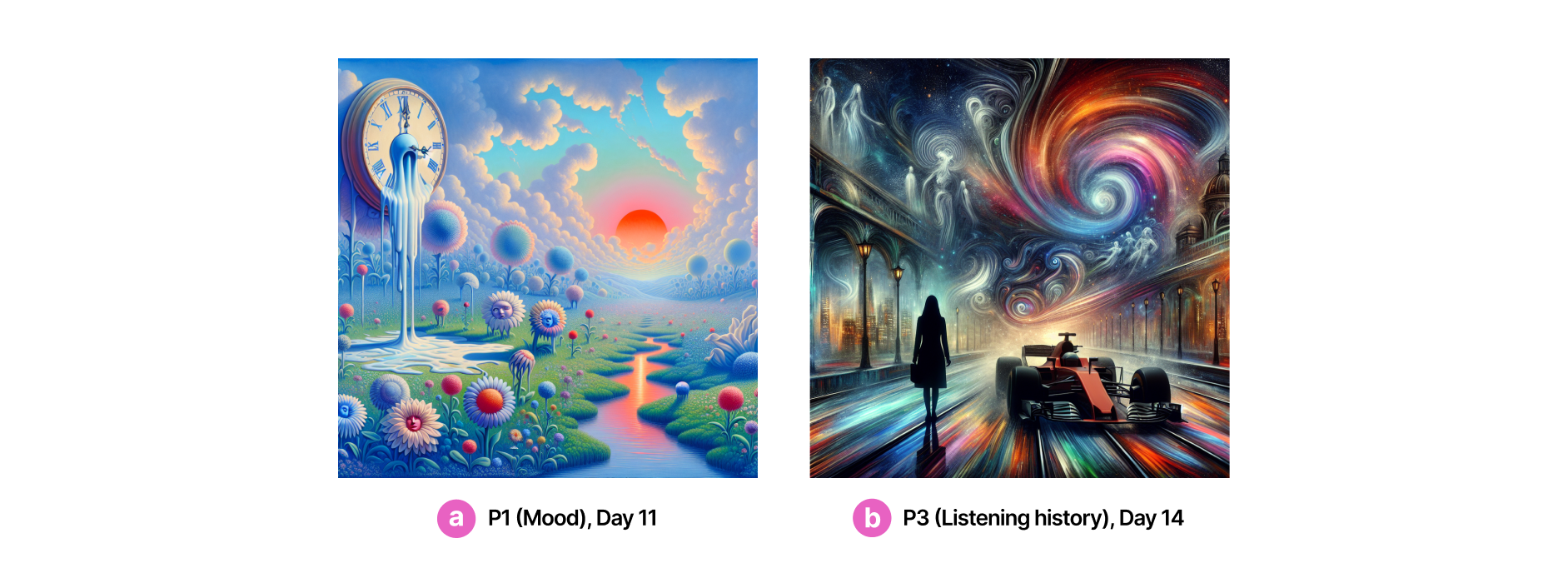}
    \caption{Image generated for (a) P1 on Day 11, (b) P3 on Day 14.}
    \label{fig:img2}
    \Description{Figure 10 illustrates two images generated by participants. (a) P1’s image on Day 11 shows a surreal landscape with a melting clock and vibrant flowers in the foreground, with a glowing sun setting over a distant horizon. (b) P3’s image on Day 14 shows a scene with a car on a street at night, a figure standing nearby, and swirling, colorful lights and cosmic patterns overhead.}
\end{figure*}

Further, participants perceived the images as a means to view their data through the lens of an external observer—in our case, the AI—offering a \textit{“third-person perspective”} (P12) and a \textit{“more objective”} (P13) interpretation of their data. In particular, P8 and P13 highlighted that this aspect differentiated the experience of image generation from taking photos or drawing:

\begin{quote}
\textit{When I draw something, it’s just my own thoughts. But what the AI created feels more objective, like how someone else might see me, and that makes it more valuable. (...) When I write my memos and journal, I can only capture the points I think of, but the AI picks up on different points and that leads to images I hadn’t thought of. That’s where I found it more surprising and fascinating.} (P13; Study hours, Interview)
\end{quote}

\begin{quote}
\textit{When I record things by just taking photos or drawing, there’s no process of that sort of reconstruction. It’s only me seeing the experience through my own lens. So I feel like I miss the opportunity to process my emotions or reflect on the experience from another person’s perspective.} (P8; Listening history (music), Interview)
\end{quote}

As a result, participants mulled over where their first-person understanding of the data aligned or diverged from the AI’s more observational and objective interpretation, which fostered introspection regarding the meaning of their personal data. Particularly, many participants (n=12) mentioned that when the generated image did not seem to match their data, it actually encouraged them to view their experiences from a new perspective. For example, P1 described how she came to reconsider her data from an unfamiliar viewpoint through an image (Figure \ref{fig:img2}-a), resulting in a new understanding of her emotions:

\begin{quote}
\textit{When I was writing, I focused only on the dark side of the event and how I felt at that time, but the image showed both the dark and positive aspects. That gave me a moment of realization, like ‘Oh, that thing wasn’t so bad after all.’ I started to see that something I thought of as entirely good or entirely bad could actually be viewed differently from another perspective, and that was something I learned through this process.} (P1; Mood, Interview)
\end{quote}

Conversely, when participants felt that the images accurately captured their thoughts or emotions, it often led them to reify and reaffirm those feelings. P3, for instance, felt that the image generated from her weekly data (Figure \ref{fig:img2}-b) precisely captured the thoughts she had that week, which helped her better understand and confirm her own thinking.

\begin{quote}
\textit{It [the image] summarized my thoughts on this week so, so, so, so well! Throughout the week, I kept thinking ‘Oh, I’ve got so much love in this world! Even if it’s not, at least the world isn’t full of weirdos…’ but I didn’t express that too directly. I think the image did a great job of capturing how I felt.} (P3; Listening history (music), Diary Day 14)
\end{quote}

Furthermore, participants engaged more deeply in the activity of reconceptualization and reinterpretation of the self by acting as a second-order observer \cite{luhmann1993deconstruction}, interpreting the AI’s interpretation in various ways. First, we observed that several participants (P3, P4, P8, P12, P14) \textit{\textbf{repeatedly generated images using the same set of data as input}}. They intentionally did so in an attempt to select the image they felt best represented their thoughts and emotions from the different variations, using this process to more clearly confirm their daily experiences, as P4 describes:

\begin{quote}
\textit{I tried to choose one specific image, thinking about which image best captures the day’s feeling, and while making that decision, I found myself reflecting on what the day’s experience was really like. That process of choosing the most fitting one really helped me look back on the day.} (P4; Step count, Interview)
\end{quote}

\begin{figure*}[ht]
    \centering
    \includegraphics[width=\textwidth]{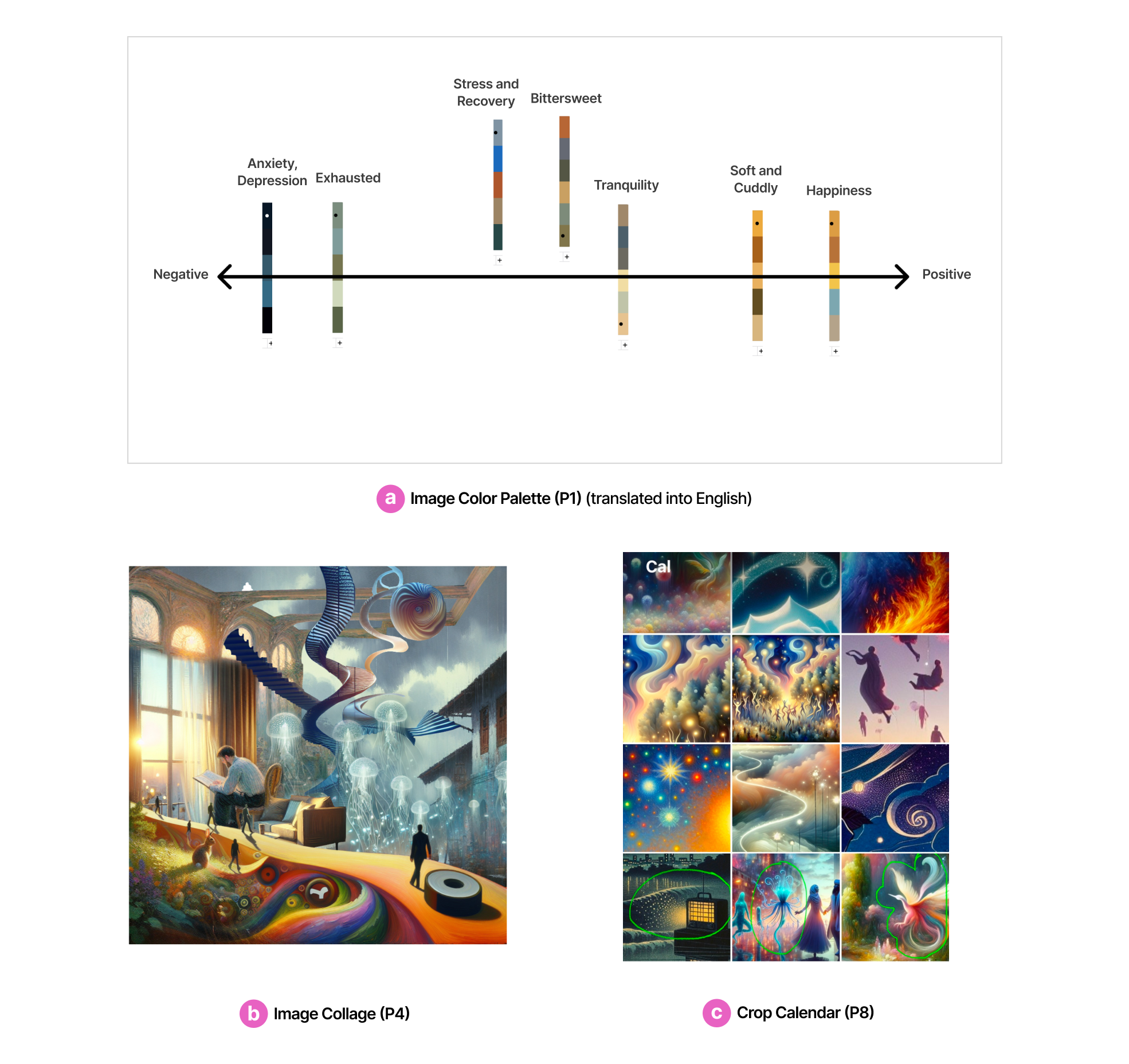}
    \caption{Examples of participants’ re-representation of generated images. (a) Image color palette created by P1. (b) Image collage created by P4. (c) Crop calendar created by P8.}
    \label{fig:img3}
    \Description{Figure 11 illustrates three examples of participants’ re-representation of generated images. (a) An image color palette created by P1. (b) Collage created by P4. (c) A crop calendar created by P8, showing a grid of cropped images.}
\end{figure*}

We also observed participants experimenting with further processing the generated images in their own ways, \textit{\textbf{transforming them into new reflective materials}}. For example, P1 created a color palette using a set of generated images (Figure \ref{fig:img3}-a). P4 made a collage using images generated over the course of a week to create a new composite image (Figure  \ref{fig:img3}-b). P8 cropped and saved specific elements from the images that caught her attention and later compiled them together (Figure  \ref{fig:img3}-c).

These practices were all aimed at better understanding how the AI had interpreted them. P1 noted, \textit{“I realized the AI was using colors to represent my emotions,”} which led her to extract colors from various images to examine how the AI expressed her feelings or memories through specific colors. P4 compared the collage she created with the weekly images, trying to uncover the differences between the elements she considered important and those \textit{the AI} emphasized. P8 also believed that the AI, through the elements in the images, was representing her emotions and daily experiences, so the process of deconstructing images was, similar to P4, also her own way of exploring the intersection between the emotions and experiences the AI captured and those that she personally found important.

Additionally, we observed that participants were interested in \textit{\textbf{interpreting the images collaboratively}} with others. They were curious about how other people would interpret the images generated from their own data and what images based on someone else’s data might look like. Several participants (n=7) actually shared their images with people around them, asking them to guess the meaning or what data might have produced the images. P12 explained that if others interpreted the images in a similar way, it would provide further validation of the meaning he had assigned to the images. P11 mentioned that he wanted to \textit{“share a new feeling”} through other people’s interpretations of the same image. As these cases show, AI-generated images point to the possibility of a new collaborative interpretation experience:

\begin{quote}
\textit{When I interpret the image by myself, I might simply guess, ‘I felt this way, and the image came out like this, so it must be expressing what I felt.’ But then I share it with someone else, we discuss it, and if that person feels something similar, I can think that my interpretation might actually be right.} (P12; Workout, Interview)
\end{quote}

Overall, participants saw the AI-generated images as reinterpretations of themselves from an external perspective, enabling new ways of understanding their data. By layering their own interpretations and even collaborating with others, they gained deeper insights and a richer understanding of themselves.

\subsection{Narrative Crafting: Shaping Personal Stories through Visual Interpretation}

We observed that participants \textit{\textbf{developed a sense of authorship}} over the AI-generated images. Although the images were generated by the AI model, the fact that those were created using participants’ own data played a significant role in shaping the sense of authorship. For example, P5 explained that she felt \textit{“as though I created it [the image]”} because the generation process involved her own data as an input. In this way, participants did not view the images as mere outputs created by AI but instead experienced a sense of creative contribution, recognizing their role as data providers who influenced and shaped the outcomes. As P7 elaborated, \textit{“I didn’t think of [the probe] just as a means. I felt like I was creating [the images] together using my heart rate and step count, and that made it feel different,”} indicating that the probe was not seen as a one-sided tool that simply produced results but as part of a co-creative process.

\begin{figure*}[ht]
    \centering
    \includegraphics[width=\textwidth]{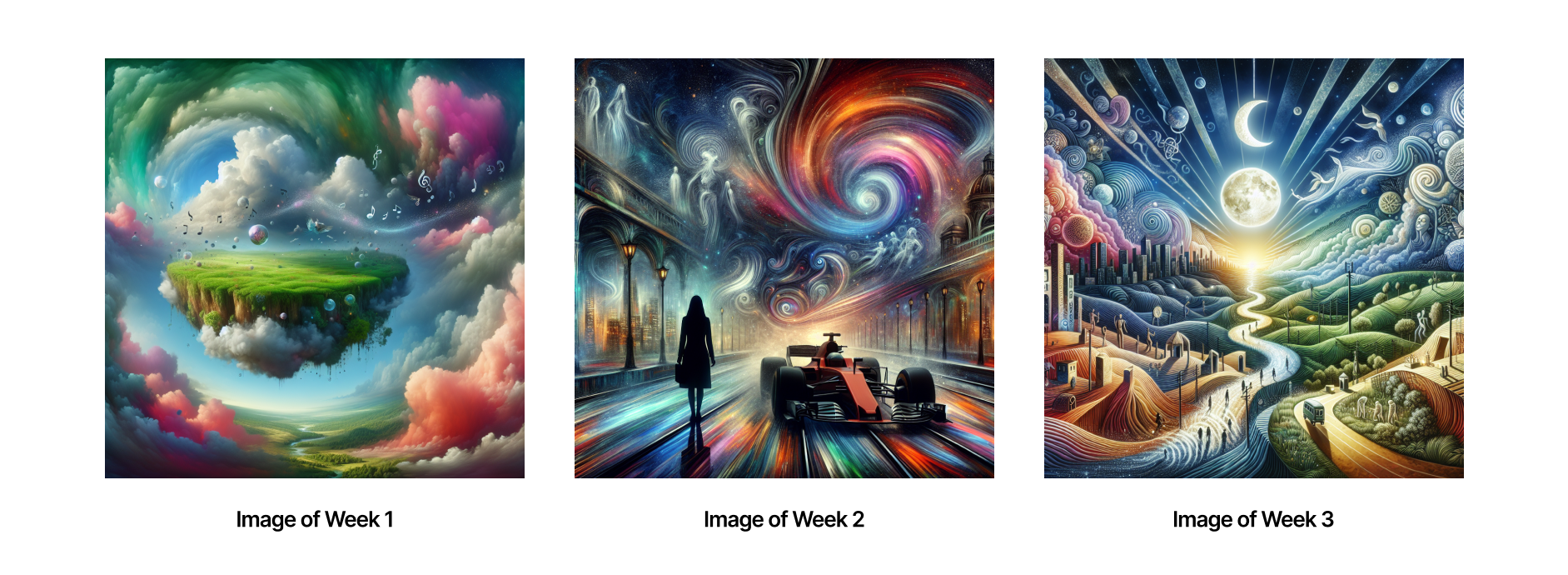}
    \caption{P3's Weekly Images.}
    \label{fig:img4}
    \Description{Figure 12 illustrates P3’s three weekly images. Image of Week 1 shows a floating island surrounded by colorful clouds and a bright sky. Image of Week 2 features a scene with a car on a street at night, a figure standing nearby, and swirling, colorful lights and cosmic patterns overhead. Image of Week 3 depicts a surreal landscape with a glowing full moon, a winding river, and radiant fields under a night sky filled with stars.}
\end{figure*}

Further, some participants \textit{\textbf{more explicitly engaged in story-making}} with their data by using unique images. Rather than treating each image separately, participants viewed the images as a part of a connected series, leading to the creation of new narratives about themselves. We observed participants tracking the flow and changes across the images over time. This was a pattern that emerged particularly after participants had collected images over an enough period of time. For instance, P3 gathered several images (Figure \ref{fig:img4}) and noticed changes among them, linking these to shifts in her own experiences and emotions, eventually weaving them into a single cohesive narrative:

\begin{quote}
\textit{You get three sets of weekly data after three weeks, and you can start writing a story. This is a bit personal, but I was going through a lot of decision-making about my academic career since mid-July. At first I was really unsure about studying abroad, but as time went on, I started to make up my mind. I didn’t explicitly write a lot about it… But when I got those three images, I could see how what was once unsettled gradually became more solid. That was really fascinating.} (P3; Listening history (music), Interview)
\end{quote}

\begin{figure*}[ht]
    \centering
    \includegraphics[width=\textwidth]{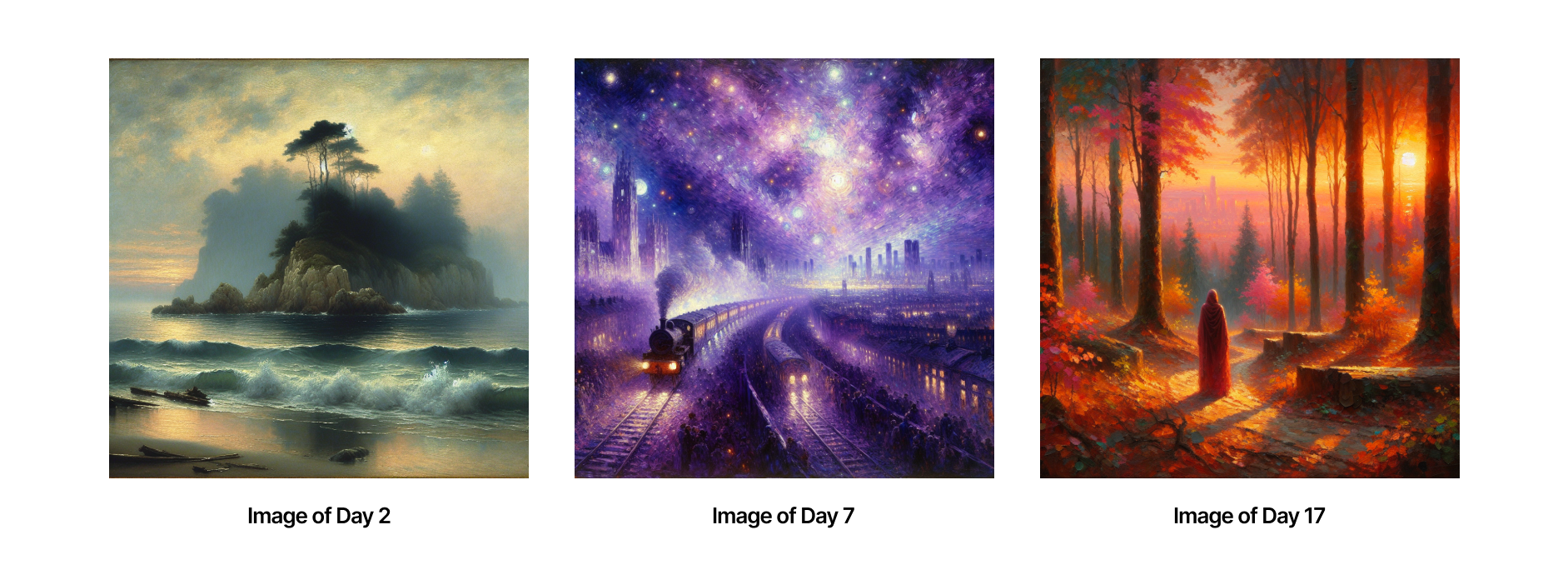}
    \caption{P3’s Images on Day 2, Day 7, and Day 17.}
    \label{fig:img5}
    \Description{Figure 13 illustrates P3’s images on Day 2, Day 7, and Day 17. Image of Day 2 shows a misty island with trees, surrounded by gentle ocean waves. Image of Day 7 depicts a dreamy cityscape with purple and blue lights, featuring a train on curved tracks. Image of Day 17 shows a person in a red cloak standing in a sunlit autumn forest, with golden light streaming through the trees.}
\end{figure*}

We also observed that, while generating images based on her listening history and the moments of listening, P3 continuously associated the images (Figure \ref{fig:img5}) with shifts in her experiences and emotions, assigning personal meaning to them. The experience of weaving these images into a coherent narrative and reflecting deeply on her emotional journey even guided P3 to gain confidence in her life decisions:

\begin{quote}
\textit{Today’s image feels like it completes the story that started with my image diaries that I’ve been keeping since mid-July. On July 15th and 16th, the images depicted Daejeon [the city’s name] as more of a new place or an island, somewhere I was traveling to, not as home. By the 21st, it’s even more so. My environment was portrayed as a new place I was visiting, not my home. But by the 31st, the image shows me looking at my old home, while the island where I’m staying now feels like home. As I’m looking at these images, and as I realized how well this system can capture my underlying intentions and emotions, I began to feel more certain about where I’m headed and where my sense of identity belongs.} (P3; Listening history (music), Diary Day 17)
\end{quote}

\begin{figure*}[ht]
    \centering
    \includegraphics[width=\textwidth]{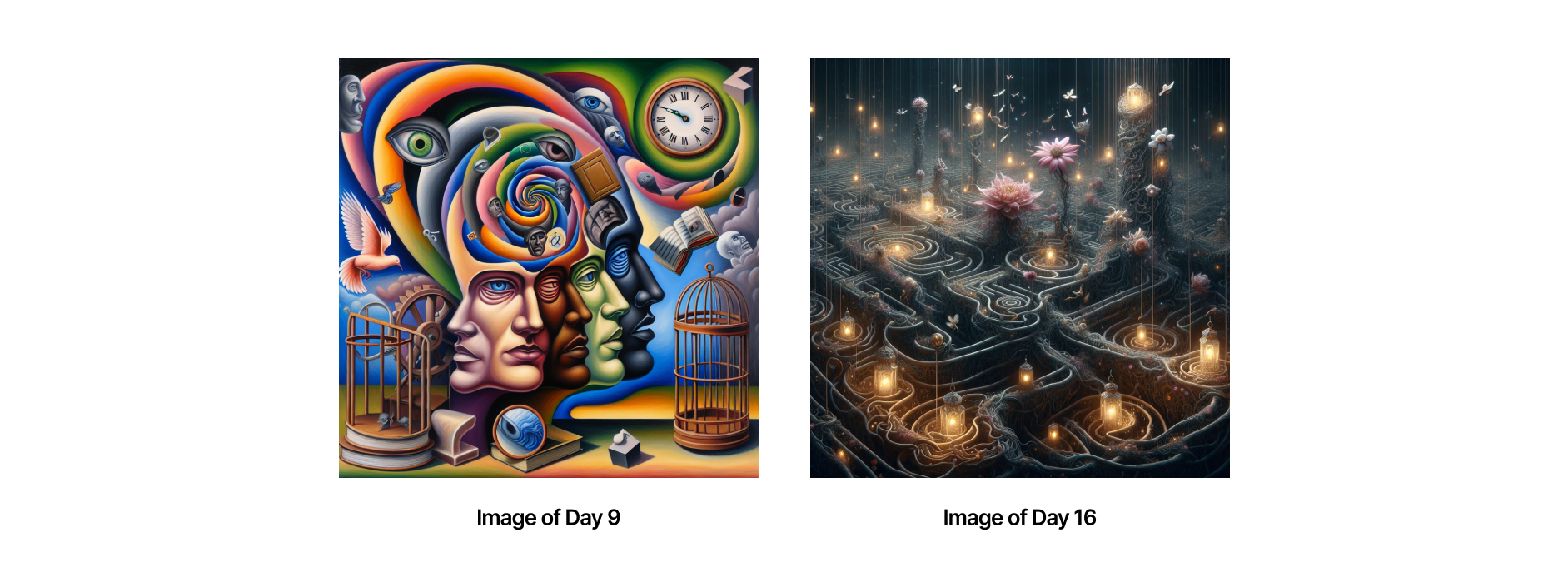}
    \caption{P6’s Image on Day 9 and Day 16.}
    \label{fig:img6}
    \Description{Figure 14 illustrates P6’s images on Day 9 and Day 16. Image of Day 9 shows a surreal composition featuring colorful overlapping faces surrounded by symbolic elements such as a clock, multiple closed birdcages, and eyes. Image of Day 16 shows a lotus pond resembling a maze, illuminated by glowing birdcages, with flying birds and blooming flowers.}
\end{figure*}

Similarly, P6 assigned meaning to recurring elements in the images and interpreted the changes in those elements as part of a narrative. Specifically, she noticed that a birdcage repeatedly appeared in the images (Figure \ref{fig:img6}) and saw the subtle changes in the cage’s form as a reflection of her evolving relationship with her screen time data:   

\begin{quote}
\textit{Last week, the cages had a more oppressive and negative vibe, but this week, there is a sense of freedom, more like an escape. I think I’m managing my app usage in a positive way, rather than just restricting it.} (P6; Screen time, Diary Day 16)
\end{quote}

Furthermore, participants expressed a desire to process the images into creative stories with more specific genres. For instance, P7 mentioned that she wanted to compile the images into a \textit{“fairy tale”} or an \textit{“essay,”} while P15 expressed interest in using multiple images to create an \textit{“animation.”} Both participants valued the serial nature in the accumulated images, believing that the flow between the images could be transformed into a cohesive story.

This, in turn, allowed us to observe that participants \textit{\textbf{regarded the images as personal belongings and felt attached}} to them. For example, P1 and P2 referred to the images as \textit{“the one and only image made from my data”} (P2) that \textit{“belongs to me”} (P1), leading them to believe that their images were unique in their origin and no one else could generate the same one. Consequently, participants developed a personal attachment to the images and a desire for ownership, suggesting that images generated based on personal data have the potential to serve as a novel form of \textit{\textbf{digital possession}} \cite{odom2010reciprocity,odom2011teenagers}. For instance, participants often saved them in their smartphone albums and used them as wallpapers on their phones or smartwatches. P11 even revisited the images saved in his smartphone album days later to recall the mood and emotions associated with the original data. Similarly, P7 expressed her motivation to save the images on her smartphone as a way to \textit{“hold onto them [the feelings evoked by the image] in some way.”} 

The above accounts collectively illustrate how participants moved beyond simply viewing the AI-generated images as outputs, instead viewing them as deeply personal creatives (co-)shaped by their own data. Through this process, participants not only developed a sense of authorship but also engaged in crafting narratives that connected the images to their evolving emotions and life experiences. These interactions revealed the unique potential of such images to serve as meaningful and personal digital possessions.

\subsection{Curiosity and Imagination-Driven Self-Tracking: Moving Beyond Task-Oriented Monitoring}

We found that AI-generated images based on personal data shifted participants away from a task-oriented mindset, leading them to a data experience \textit{\textbf{driven by curiosity and imagination}}. First, we noticed that participants considered \textit{\textbf{the anticipation of what kinds of images would be generated}} to be a more meaningful part of the data collection process, rather than focusing on whether they had achieved their goals. Many participants were excited about the idea of new images being generated from their data, and this curiosity added enjoyment to the data collection process, serving as a source of motivation for consistent tracking. For instance, P4 mentioned that she rarely checked her step count often or would look at it all at once, but the anticipation of what kind of image would be generated made her want to check her tracking data more frequently. Also, P7 mentioned tracking data with \textit{“a sense of excitement about what image would come out every day”}:

\begin{quote}
\textit{I was really looking forward to seeing what image would come out that day. When I used another health app, there was no such sense of anticipation at all. It just tracked the data, and that was all. Nothing came out of it. But with this app, there was always this expectation, an output that I can gain.} (P7; Health, Interview)
\end{quote}

\begin{figure*}[ht]
    \centering
    \includegraphics[width=\textwidth]{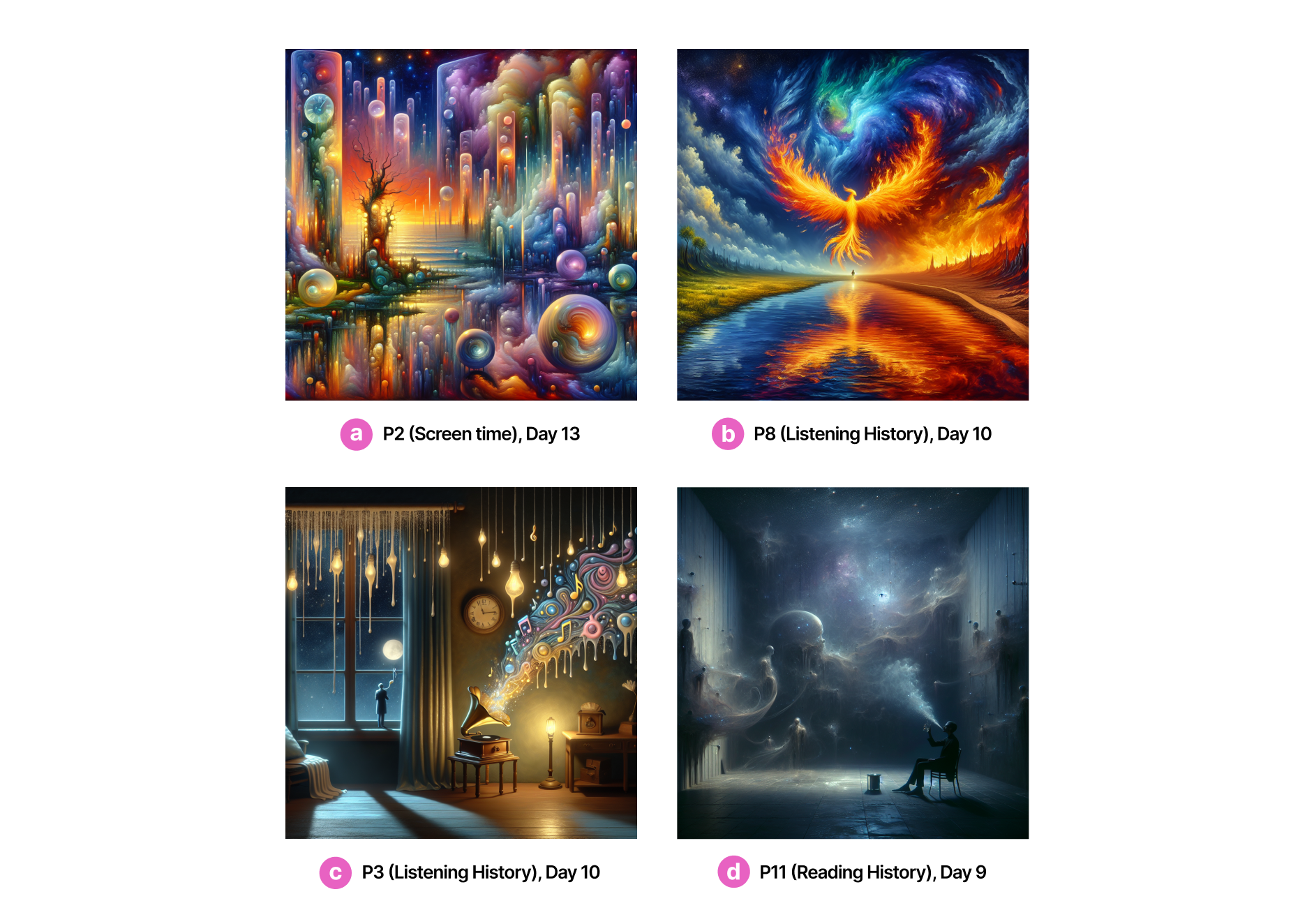}
    \caption{(a) P2’s Image on Day 19, (b) P8’s Image on Day 10, (c) P3’s Image on Day 10, (d) P11’s Image on Day 9.}
    \label{fig:img7}
    \Description{Figure 15 illustrates four images generated by participants. (a) P2’s image on Day 13 shows a surreal scene of a river where vibrant colors and clouds seem to melt and blend into the water, creating a flowing, dreamlike effect. (b) P8’s image on Day 10 shows a vibrant phoenix rising from its reflection in a calm lake, surrounded by fiery skies and dramatic clouds. (c) P3’s image on Day 10 shows a cozy indoor scene with a gramophone playing music, its melodies visualized as swirling, colorful patterns floating into the air. (d) P11’s image on Day 9 shows a dark scene of a person seated in a vast, cosmic space-like room, with ethereal mist blending seamlessly into the surroundings.}
\end{figure*}

We also found that the way AI-generated images represented data helped participants \textit{\textbf{alleviate their obsession and feelings of guilt}} resulting from numerical forms of tracking data, facilitating a mindful tracking experience. This was especially evident among participants (P2, P6, P13) who used the probe for the purpose of time management. In the process of interpreting a blurry association between the image and the source data, they were able to reflect on their compulsions and obsessions from a new perspective, shifting the focus of attention from the figures. For example, by viewing images (e.g., Figure \ref{fig:img7}-a), P2 became less \textit{“guilty”} about her screen time as she reflected on the actual experiences that the numbers entailed:

\begin{quote}
\textit{Sometimes the images would show time beautifully melting away, and I wondered if it tried to say my time is slipping by. But then, I’d also see the images connected to positive aspects of internet use, like telling me that exploring various world wide webs on my phone isn’t always a bad thing. I used to feel really guilty about using my phone, but when those kinds of images appeared, it made me think I don’t need to be so negative about it.} (P2; Screen time, Interview)
\end{quote}

P6 also thought the images might be signaling that she was overly controlling her screen time, attributing this to the ambiguity of the image. This eventually led to a shift in her perspective on screen time data; she mentioned that her original goal of tracking screen time was to \textit{“live a more self-directed life,”} but the images made her think that \textit{“perhaps I’m too obsessed with this.}”

Also, the images encouraged participants to \textit{\textbf{make small changes in their daily lives}}. P4 mentioned feeling curious about \textit{“what  it [the generation model] would say this time”} when she intentionally altered her step count, saying that the process was \textit{“like a game.”} P8, curious about how a different listening history would influence image generation, decided to listen to K-pop dance tracks—something she hardly enjoyed, eventually reflecting on her day through the newly generated image (Figure \ref{fig:img7}-b):

\begin{quote}
\textit{Looking back at my previous history, I noticed a pattern in music preferences and the corresponding images. Today, I wanted to see something different from the usual nature-inspired images that reflected mystical love and the passage of time. So I chose fast-tempo K-pop dance songs, which was a bit outside of my usual pattern. I was overwhelmed with work, I felt like listening to something upbeat, and I was curious to see if a new kind of image would be generated. It was good that the images generated were just as fresh as I had hoped.} (P8; Listening history (music), Diary Day 10)
\end{quote}

Lastly, participants expressed a desire to use the images as a medium for \textit{\textbf{genuine self-expression}}. Participants (P1, P3, P9, P10, P11, P14) anticipated that the images could be used to share their emotions or situations without a sense of self-promotion that provokes comparison and competition, since the images didn’t reveal the associated data directly. As P14 described, the images \textit{“contain my data but with a certain level of ambiguity,”} making  them suitable for sharing when \textit{“you just want to convey what you were feeling or what the situation was like at the time.”} For instance, P13 mentioned that if he shared the image generated from his study hours instead of the raw numbers, the focus of sharing would be on personal thoughts and emotions, not on fueling competition or comparison based on the amount of time studied:

\begin{quote}
\textit{As with apps like Yeolpumta [the name of a study hour tracking app], seeing others’ study hours makes you want to compare yourself with them, but with these images, it’s less about comparison and more about seeing what someone felt about, their emotions related to the subject they studied.} (P13; Study hours, Interview)
\end{quote}

P3 and P11 even uploaded the images generated from their data (e.g., Figure \ref{fig:img7}-c, Figure \ref{fig:img7}-d) to Instagram to express their emotions, noting that posting these images allowed them to share feelings tied to the data with other users while keeping explicit details private:

\begin{quote}
\textit{[The image] shows my emotions, but someone who doesn’t know the actual context has a lot of room for interpretation… And that’s part of why people use Instagram, right? Everyone wants a bit of attention, and the AI finds a sweet spot so well. It might be an image that looks sad but also beautiful, and while it seems like there’s a deeper story, and [the viewer] doesn’t know that. That kind of ambiguity strikes the balance between the desire to share a bit of myself and avoid being overly exposed. }(P3; Listening history (music), Interview)
\end{quote}

These findings collectively show that AI-generated images shifted participants from task-oriented data tracking to more curiosity-driven, imaginative experiences. The ambiguity of the images encouraged self-reflection, reducing guilt or obsession with numerical data and fostering a more mindful approach. Participants also saw the images as a safe, creative way to express emotions without the pressure of comparison.

\section{Discussion}
Our findings demonstrate that AI-generated images based on personal data immerse individuals in the process of interpreting images created by AI’s autonomy, leading to new meaning-making experiences with their data. Building on these findings, we highlight the potential of generative AI in facilitating deeper self-reflection through the co-interpretation of personal data (§ 5.1) and propose possible domains and designs where AI-generated images can be applied as a new opportunity (§ 5.2). Finally, we discuss the inherent uncertainty of AI-generated outputs as a double-edged sword that may present a unique challenge posed by generative AI (§ 5.3). 

\subsection{Deepening Self-Reflection through Co-Interpretation of Personal Data with Generative AI}

Throughout our study, we observed that meaning-making through generative AI occurred as a reciprocal process where AI interpreted human data, and humans then reinterpreted the AI’s output, fostering a deeper experience of self-reflection. We notice that this aligns with the concept of \textit{\textbf{co-interpretation}}  \cite{romero2005preliminary,romero2008alien,pousman2008living}, a process in which both the user and the machine assign meaning to an artifact, which can \textit{“open unusual viewpoints onto everyday human activity, create pleasure, and provide opportunities for contemplation and wonder”}  \cite{romero2005preliminary}. This interplay between human and AI interpretation transformed the typically passive role of data consumption into an active, reflective engagement, where meaning was continuously constructed and reconstructed in a dynamic, evolving dialogue between the participant and the AI.

In particular, we observed that co-interpretation, a concept defined before the advent of LLM-based generative models, can be further facilitated by the unique features of emerging generative AI. This led to the realization that co-interpretation now occurs in novel ways, distinct from previous approaches. For example, due to its generative variability \cite{weisz2023toward,weisz2024design,muller2024genaichi}, AI produces variable and uncertain generative outcomes, allowing interpretation from multiple perspectives that make more room for defamiliarization \cite{bell2005making} of the original data. Also, even when using the same input data, the generation of multiple outputs provoked participants to explore the intersection between human interpretation and AI interpretation, facilitating a meaning-making process where they could view and better understand their data from new angles. Plus, the complete black-box nature of image generation made the images even more ambiguous and open to interpretation \cite{gaver2003ambiguity,sengers2006staying}. As a result, participants engaged in a process of guessing the meaning of the images and speculating on how the AI interpreted their data and how this was reflected in various attributes of the image (i.e., elements, color, mood), in which they were able to form their own personal meanings.

Therefore, we propose the need for designing generative AI that produces images with a space for co-interpretation to foster meaning-making from personal data, rather than images that serve merely as a ‘right answer’ that most accurately depicts a situation or emotion. This suggests the possibility of an alternative design approach for human-AI interaction, distinct from currently dominant task-oriented methods used for AI-driven reflection, such as providing accurate analysis, prediction, or guesses of emotions and past experiences \cite{g2021reflection,hollis2017does,rohani2020recommending}. Instead of focusing on how to design an AI system that strives to achieve a desired outcome, we highlight the potential for a new interaction design that centers on the human reinterpretation of the AI’s outputs in a user’s own way. This aligns with Brand et al.’s works \cite{brand2021design,brand2023envisioning}, and we aim to build on this line of research. Specifically, we extend this work by demonstrating the potential of image-generative AI for reflection, emphasizing how co-interpretation—where human interpretation is layered onto AI-generated outputs—played a unique role in this process. We also aim to reiterate the value of AI as a reflective partner \cite{van2023objective,van2022ceci}, by introducing the concept of co-interpretation between personal data, the end-user, and AI, thus expanding the scope of how AI can be integrated into reflective and meaning-making processes. Also, Sarkar \cite{sarkar2024ai} has proposed a novel perspective on positioning AI as a \textit{provocateur}, emphasizing its potential to stimulate critical thinking. Our work resonates with this direction by demonstrating how generative AI can enable active meaning-making through co-interpretation, allowing users to move beyond the passive consumption of AI-generated outputs.

From this perspective, we can imagine a new form of human-AI interaction that embodies the experience of co-interpretation. For instance, as our findings suggest, to facilitate co-interpretation, we can design AI deliberately not to provide clear, explicit explanations on the prediction results, and rather make the system \textit{\textbf{stimulate a user’s curiosity and leave room for his or her interpretation}} on the results. Also, just as our study participants found interpretive intersections between themselves and the AI through various distinct outputs, the system can be designed to help users engage in \textit{\textbf{deeper reflection from multiple perspectives}} on their own data. One key aspect to consider in this design, as highlighted by Romero and Mateas \cite{romero2005preliminary}, is finding the right balance; AI-generated outputs should neither be too obvious that they feel like a one-to-one mapping to the data, nor be overly complex to the point of seeming random. Striking this balance will be crucial to effectively foster meaningful interpretation experiences. To achieve this balance, a potential approach would be to design a system that, as shown in our findings, allows users to reinterpret AI’s outputs in new ways by \textit{\textbf{providing hints that encourage them to explore the AI’s interpretation}}, such as giving users the choice to check out the prompts used for generation. Such a way of offering subtle clues or cues would help users engage in the co-interpretation process without overwhelming or limiting their creative reflection.

\subsection{AI-Generated Images as an Interpretive Medium}
Throughout our study, we observed that AI-generated images based on personal data served as an ‘interpretive’ medium by inviting participants to actively make personal meaning from their data. We now propose various possible domains as a new opportunity to leverage the value of such an interpretive medium.

\subsubsection{Self-Tracking}
Prior studies in the self-tracking domain have suggested that when data visualizations emphasize unmet goals and failures, they can trigger negative thought cycles \cite{niess2020exploring,spiel2018fitter}. Strömel et al. \cite{stromel2024narrating} emphasize that preventing such rumination is a critical design consideration for systems supporting reflection. We propose that when AI-generated images of personal data are integrated into self-tracking tools, they can offer a new personal informatics experience, where users reflect on their emotions rather than engage in self-criticism resulting from cold-hard figures and graphs. Plus, the fun of anticipating AI-generated images may inspire a more positive engagement with personal informatics. Ultimately, AI-generated images offer the potential to transform self-tracking tools by shifting the focus from goal-oriented evaluation to a more exploratory and self-reflective interaction \cite{choe2017understanding,baumer2014reviewing,baumer2015reflective,jung2020search} with personal data.

\subsubsection{Social Media}
Supporting indirect and ambiguous self-disclosure while protecting users’ privacy and sensitive information on social media has been a key topic in this domain \cite{andalibi2018testing,andalibi2017sensitive}. Echoing previous discussions \cite{hua2024generative}, our work reveals that personal-data-driven images hold value as a new social medium for sharing their thoughts and emotions in less direct and more interpretive ways, opening up new avenues for self-expression that balance personal privacy with meaningful communication. This potential was highlighted in our findings (§4.4), in which we observed several participants uploading AI-generated images to Instagram to convey their emotions associated with the data without revealing explicit details. Moreover, these images hold the potential to support emotional communication in an indirect manner \cite{brave1997intouch,gaver2023living,strong1996feather}, opening up a new opportunity for more nuanced expressions of feelings that foster new types of social interactions \cite{elsden2016metadating}.

\subsubsection{Journaling and Life-logging}
Our findings show that participants reminisced about memories and emotions associated with their data through the AI-generated images, indicating life-logging as another potential application domain. In prior cases of systems supporting journaling, AI has been used to generate reflective questions \cite{kim2024mindfuldiary} or curate personal data as a resource \cite{apple}. Building on these ideas, our findings show the possibility of AI-generated images using personal data as fresh material for journaling. For instance, instead of simply suggesting “music you listened to a month ago” as diary materials, “an image generated from the most frequently played music over the past month” could be presented, encouraging users to reflect not only on the music itself but also on the emotions they experienced while listening to it. Further, images as new representations of personal data can act as a new form of \textit{digital memento} \cite{dib2010sonic,petrelli2010family} or as a medium for \textit{documentary tracking} \cite{elsden2017designing}, making it easier for users to log their life stories in a creative and engaging way.

\subsubsection{Digital Possession}
Our findings reveal that personal data-driven images can serve as a novel form of digital possession, offering new opportunities for designing systems that integrate such possessions into everyday life. This potential was most prominently demonstrated in §4.3, where participants developed a sense of authorship over the images and crafted personal stories through their interpretations. It was further supported by participants’ desire to save or print images for journaling (§4.1) and their attempts at self-expression through these images (§4.4)—behaviors that highlight their desire to own and utilize these images as personally meaningful possessions. Aligning with previous works, which has shown that people construct new value as digital possessions by developing personal narratives around media content (e.g., music, books, photos) \cite{odom2011teenagers, odom2014placelessness, watkins2015digital, watkins2012attachment,belk2016extended}, our findings demonstrate that AI-generated images from personal data can also become objects that people collect, cherish, imbue with meaning, and seek to preserve \cite{golsteijn2012towards}. In particular, we expect that the ‘shared agency’ \cite{elsden2016s, gulotta2015curatorial} between generative AI and users throughout the process of co-interpretation would provide unique experiences of such a new type of digital possession, by fostering a collaborative creation, curation, and preservation process as well as enriching the sense of ownership.

In light of this, we suggest future research into longer-term investigations into AI-generated images as digital possessions. While the significance of our study lies in uncovering the potential of this new type of digital possession, longer-term studies are expected to illuminate more nuanced user experiences of it \cite{odom2014designing, odom2015understanding,odom2019investigating}. A future research agenda might include exploring how the accumulation of images over time could enable people to craft stories from these larger collections, how the integration of metadata adds personal and social meaning, or how transitions between material and virtual forms occur and facilitate reflection \cite{odom2014placelessness, odom2010virtual, odom2011teenagers}.

\subsubsection{Mental Wellbeing}
Our findings demonstrate that AI-generated images not only help people simply revisit past emotions, but also allow for multiple interpretations of those emotions (i.e., \cite{staahl2014evocative}). Therefore, we see the potential for using such images in the systems for mental wellbeing to support people’s self-observation and emotional understanding. For example, similar to the practice where individuals use image prompts to assist in self-care by visualizing their feelings \cite{capel2024studying}, we can imagine a system where individuals who struggle to articulate their emotions in words could input descriptive adjectives and explore the generated images. This process would help them find an image that best resonates with their emotions and encourage them to express those feelings in their own words. Also, inspired by the interactive game \textit{Say what you see!} \cite{saywhat}, we can envision a feature that helps users guess the AI’s interpretation by articulating their emotions when viewing the generated images, and then comparing their interpretations with the AI’s. However, we note that designers should be aware of potential risks related to these approaches, such as the danger of evoking negative emotional experiences, and it will be crucial to incorporate features that prevent or mitigate such negative experiences when designing systems for mental wellbeing. Building on this, we discuss more concerns related to generating images using personal data as input in the following section.

\subsection{Output Uncertainty as a Double-Edged Sword}
As reported in our findings (§4.1), a few participants reported unfavorable experiences with the images when the generated images mirrored their negative emotions, thereby reinforcing those feelings. While the inherent unpredictability and uncertainty of AI-generated images has shown its value in providing an unfamiliar perspective on the original data, it also highlighted the risk of triggering unpleasant emotional responses. 

To mitigate these risks, it is essential to carefully embody principles supporting mental wellbeing when designing systems. First, designers should carefully choose AI models that prioritize preventing the creation of harmful content, such as violent, adult, or hateful imagery \cite{dallepolicy}, which could evoke negative emotions. Additionally, designers might consider choosing models that minimize the inclusion of commonly disliked, undesirable elements (e.g., avoiding images that fall into the ‘uncanny valley’). Furthermore, fine-tuning models to produce images in a tone that avoids excessive emotional stimulation \cite{lin2024make, yang2024emogen} and supports emotion regulation \cite{gross2015emotion} could be another approach. However, due to the inherent unpredictability of generative AI outputs, it is necessary to adopt a more designerly approach for enabling users to experience the images in an emotionally safe manner. As Cognitive Load Theory suggests that cognitive overload can increase emotional stress and fatigue \cite{sweller2011cognitive}, introducing pauses during the image generation process may help prevent users’ emotional overload. For example, the system could ask users’ current emotional state and provide options for proceeding with the display of the final image. Since users may have different emotional triggers, providing customizable features could allow them to tailor the system to their emotional readiness \cite{zhang2021designing}. For example, the system could allow users to set the emotional tone of the image before presenting the generated images, while still ensuring the serendipity of outputs that can inspire creativity and reflection. Furthermore, in line with Cognitive Behavioral Therapy principles that highlight practicing emotional awareness \cite{hofmann2012efficacy}, the system could include features that ask reflective questions to help users regulate emotions provoked by the images. By embodying these principles to enhance emotional wellbeing, we expect that designers can reduce the risk of evoking negative emotions through AI-generated images. These are just a few initial examples, and we encourage further research in this area.

Meanwhile, we also propose there may be another more nuanced approach to addressing such ‘negative’ emotions. As conceptualized in the notion of \textit{uncomfortable interaction} \cite{benford2012uncomfortable}, deliberately incorporating elements of discomfort can offer users benefits such as entertainment, enlightenment, and sociality. In a similar vein, Bopp et al. \cite{bopp2016negative} demonstrated that negatively valenced emotions (e.g., sadness) can contribute to positive experiences, underscoring that such emotions do not necessarily render the experience itself negative but instead foster thought-provoking engagements. Benford et al. \cite{benford2012uncomfortable} further proposed leveraging emotions arising from unpredictable control and surprising system responses as a tactic for designing uncomfortable interactions, which resonates with the value of inherent unpredictability and uncertainty of AI-generated outputs highlighted in our findings. Thus, potential negative emotions evoked by AI-generated images may not always be considered as risks to mitigate; instead, they might also be opportunities to provide users with meaningful reflective \cite{fleck2010reflecting,mols2016technologies,baumer2015reflective}, introspective experiences, echoing the findings of Brand et al.’s work \cite{brand2023envisioning} where participants valued the confrontational qualities of AI outputs despite experiencing discomfort, underscoring AI’s role in fostering critical self-confrontation. We therefore invite future works to explore how the deliberate incorporation of discomfort and unpredictability in AI-generated images can be harnessed to create deeper, more reflective interactions with personal data, striking a balance between emotional challenges and meaningful engagement.

\section{Limitations and Future Work}

We acknowledge several limitations in our study and suggest directions for future research.

First, our study involved participants exclusively from South Korea. Expanding future studies to include participants from diverse cultural backgrounds could provide a broader understanding of how cultural factors influence the meaning-making of AI-generated images.

Second, while the 21-day study was effective in exploring new opportunities for image-generative AI in personal data meaning-making, this research also uncovered promising new avenues for future research that could be meaningful to explore through longer-term studies (e.g., §4.3). For example, as mentioned in §5.2.4, longer-term studies could delve into deeper experiences with AI-generated images derived from personal data as digital possessions. Also, researchers might examine the behavioral changes driven by integrating image-generative AI into self-tracking tools, or explore the in-depth interpersonal dynamics that arise from socially sharing such images. This presents directions for future longer-term studies, allowing researchers to explore the sustained impact of image-generative AI on personal data interpretation across various dimensions.

Third, we employed an autobiographical design approach to mitigate potential risks, such as privacy concerns, before involving external participants. While we carefully addressed these concerns in the design of the probe by selecting models and establishing image-generation rules, further research is needed to explore how such risks can be more thoroughly addressed. Future studies will be necessary to develop detailed design guidelines that effectively address and mitigate these concerns.

Meanwhile, our study primarily focused on the visual modality, leveraging generative AI’s reflective potential through image generation. However, generative AI technologies are now capable of producing various outputs, such as sound, video, and other modalities. Exploring how personal data, when transformed into various forms or multi-modalities, affect user experiences could be an interesting avenue for future research.

\section{Conclusion}
Our work investigated how AI-generated images based on personal data present new design opportunities for meaning-making. We conducted a 21-day diary study and interviews with 16 participants, informed by an autobiographical-design-based formative study. Our findings show that AI-generated images acted as a unique interpretive medium, encouraging participants in a new way of reflective meaning-making with their data. We highlight the potential of image-generative AI in deepening self-reflection on personal data through co-interpretation, where participants reconstruct the meaning of personal data through an additional layer of machine interpretation. We hope that our work contributes to HCI and design research by articulating and deepening our understanding of generative AI as a material for reflective design.

\begin{acks}
We sincerely thank all participants in our study. We are also grateful to the anonymous reviewers for their insightful comments and suggestions for this paper. This work was supported by the National Research Foundation of Korea (NRF) grant funded by the Korea government (MSIT) (No. NRF-2021R1A2C2004263).

\end{acks}

\bibliographystyle{ACM-Reference-Format}
\bibliography{Reference}

\clearpage
\appendix

\section{Prompt for Image Generation}
\textbf{(a) Initial Setting}\\
\textbf{System Prompt: }You are an assistant who creates image prompts. Your goal is to support self-reflection by creating an image that allows users to reflect on their data in a subjective way. You will be provided with the personal data related to “\textit{[DATA TYPE]} of a day”. Given the user’s \textit{[DATA TYPE]} data, your task is to create an original and unexpected image prompt for DALL-E to produce an image that is inspired by your interpretation of this data. You MUST follow the provided [Image Generation Rules] when generating image prompts. Provide your output in an image prompt format.
\\
\\
\textbf{(b) Image Generation Rules}\\
\textbf{[Image Generation Rules]}\\
\textbf{RULE 1. }Extract keywords based on the interpretation of the data before generating the image. Keywords can include scenes, objects, landscapes, colors, patterns, moods, etc. The image should be generated based on these keywords.\\
\textbf{RULE 2. }When creating keywords, reflect the interpretation of the data rather than the type of data itself. For example, interpret “24-minute exercise” creatively instead of simply noting “exercise”. The image should embody these keywords in a creative manner. Do not draw objects that directly depict the data type (e.g., workout, screentime, etc) and “personal data” itself.\\
\textbf{RULE 3. }Do not include any numbers, letters, or text in the image.\\
\textbf{RULE 4. }Generate an image that is not easily associated with the data provided, ensuring originality and unexpectedness.\\
\textbf{RULE 5. }Avoid repetitive image prompts by using diverse keywords and styles. The subject, mood, texture, and style of the image should vary to maintain diversity and creativity.\\
\textbf{RULE 6. }Draw in one of the following styles. Use a different style for each image to ensure a wide variety:\\
- Medium: painting, photo, sketch, cartoon, icon, vector, graffiti, 3D render\\
- West Figurative Premodern: Baroque, High Renaissance, Impressionism, Medieval, Pointillism, Neoclassicism\\
- West Figurative Modern: Pop Art, Surrealism, documentary photography, Art deco, Hippie movement, photorealism\\
- Non-West Abstract Modern: Ukiyo-e, Chinese ink wash painting, Kerala mural, Mayan art, African masks, ancient Egyptian art, thangka\\
- Non-West Abstract Modern: Mola art, Geometric Islamic art, Mexican Otomi, Andean textile, Aboriginal art\\
- West Abstract Modern: action painting, Op art, Bauhaus, Cubism, Dadaism, Futurism\\
\textbf{RULE 7. }Response should only be the image prompt, do not give any prefix such as “prompt”, “image prompt”, or “prompts for DALL-E” and do not use double quotes at the start and end of the response.
\\
\\
\textbf{(c) Processor of User Input Data}\\
\textbf{User: }\\
\textit{[DATA TYPE]} of a day:\\
\textit{[INPUT FIELD OF DATA TYPE]: <user’s input>}
\\
\\
(Example)\\
Habit Tracking of a day:\\
- Habit: Morning Meditation\\
- Rating: 4/5\\
- Reason: Maintained concentration for most of the session, with two brief distractions.

\end{document}